\documentclass[letterpaper,11pt]{article}

\usepackage[utf8]{inputenc}
\usepackage[T1]{fontenc}
\usepackage{cfr-lm}

\usepackage{relsize}
\usepackage{exscale}

\usepackage{stmaryrd}
\usepackage{cite}
\usepackage[dvips]{graphicx}
\usepackage{graphpap}
\usepackage{ifthen}
\usepackage{enumerate}
\usepackage{amsmath}
\usepackage{amssymb}
\usepackage{mathrsfs}
\usepackage[errorshow]{tracefnt}
\usepackage{fancybox}
\usepackage{amsfonts}
\usepackage{amssymb}
\usepackage{multirow}
\usepackage{array}
\usepackage{mathtools}
\usepackage{soul}
\usepackage{pict2e}
\usepackage{bbold}
\DeclareMathAlphabet{\mathpzc}{OT1}{pzc}{m}{it}

\usepackage{lscape}
\usepackage{xypic}
\usepackage[dvipsnames]{xcolor}
\usepackage[
color,matrix,arrow]{xy}
\usepackage{setspace}

\usepackage{epsf}

\usepackage{enumitem}

\usepackage{graphicx}

\renewcommand{\leq}{\leqslant}
\renewcommand{\geq}{\geqslant}

\makeatletter
\newcommand{\thickhline}{%
    \noalign {\ifnum 0=`}\fi \hrule height 1pt
    \futurelet \reserved@a \@xhline
}
\newcolumntype{'}{@{\hskip\tabcolsep\vrule width 1pt\hskip\tabcolsep}}
\makeatother
\newcolumntype{"}{@{\hskip\tabcolsep\vrule width 1.5pt\hskip\tabcolsep}}
\makeatother

\newcommand{\scr}{\mathscr}

\def\ie{{\it i.e.}}

\def\apriori{{\it a priori}}
\def\eg{{\it e.g.}}

\newcommand\BB{{\scr B}}

\def\boxit#1{\vbox{\hrule\hbox{\vrule\kern3pt
             \vbox{\kern3pt#1\kern3pt}\kern3pt\vrule}\hrule}}

\newcommand{\Blue}[1]{{\color{blue} #1}}

\newcommand{\beq}{\begin{equation}}
\newcommand{\beqn}{\begin{equation*}}
\newcommand{\eeq}{\end{equation}}
\newcommand{\eeqn}{\end{equation*}}
\newcommand{\beqa}{\begin{eqnarray}}
\newcommand{\beqan}{\begin{eqnarray*}}
\newcommand{\eeqa}{\end{eqnarray}}
\newcommand{\eeqan}{\end{eqnarray*}}
\newcommand{\bdm}{\begin{displaymath}}
\newcommand{\edm}{\end{displaymath}}

\newcommand{\wt}{\widetilde}

\newcommand{\ba}{\begin{array}}
\newcommand{\ea}{\end{array}}

\newcommand\nn{\nonumber}

\newcommand\benu{\begin{enumerate}}
\newcommand\eenu{\end{enumerate}}
\newcommand\bit{\begin{itemize}}
\newcommand\eit{\end{itemize}}

\def\tr{\mathrm{tr\,}}
\def\dim{\mathrm{dim\,}}

\def\der'{\mathfrak{der}'\,}
\def\der{\mathfrak{der}\,}
\def\str'{\mathfrak{str}'\,}
\def\str{\mathfrak{str}\,}

\newcommand{\de}{\delta}

\newcommand{\ad}{\mathrm{ad}\,}

\def\fg{{\mathfrak g}}

\def\fgrplus{{\mathfrak g}^+}
\def\sB{{\scr B}}

\def\sh{\sharp}
\def\fl{\flat}

\def\*{\partial}

\def\tk{\widetilde k}
\def\tR{\widetilde R}
\def\ttR{\widetilde{\widetilde{R\,}}\!}

\def\id
{\mathbb{1}}

\def\cupp{\mathrel{\mathsmaller\bigcup}}

\def\adj{\hbox{\bf adj}}
\def\mult{\hbox{mult}}

\def\Weight#1#2#3#4#5#6#7#8{\bigl({}_{#1#2#3#4}^{\mathstrut}{}_{#5}^{#8\mathstrut}{}_{#6#7}^{\mathstrut}\bigr)}

\def\Root#1#2#3#4#5#6#7#8{\bigl[{}_{#1#2#3#4}^{\mathstrut}{}_{#5}^{#8\mathstrut}{}_{#6#7}^{\mathstrut}\bigr]}

\def\DPPWeight#1#2#3#4#5#6#7#8#9{\bigl({}_{#1}^{\mathstrut}{}_{#2}^{#8\mathstrut}{}_{#3#4}^{\mathstrut}{}_{#5}^{#9\mathstrut}{}_{#6#7}^{\mathstrut}\bigr)}

\def\rep#1{\mathbf{#1}}

\def\PP{{\mathbb P}}

\def\nullrep{\{0\}}

\RequirePackage{hyperref}

\numberwithin{equation}{section}

\begin{document}


\frenchspacing

\null\vspace{-18mm}

\includegraphics[width=4cm]{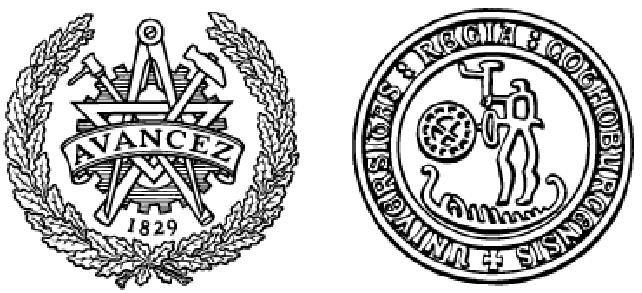}

\vspace{-17mm}
{\flushright Gothenburg preprint \\ August, 2019\\} 

\vspace{4mm}

\hrule

\vspace{8mm}


\pagestyle{empty}

\begin{center}
  {\Large \bf \sc Tensor hierarchy algebras and extended geometry I:}
  \\[3mm]
  {\large \bf \sc Construction of the algebra}
    \\[8mm]
    
{\large
Martin Cederwall${}^1$ and Jakob Palmkvist${}^{1,2}$}

\vspace{8mm}
       ${}^1${\it Division for Theoretical Physics, Department of Physics,\\
         Chalmers University of Technology,
 SE-412 96 Gothenburg, Sweden}

\vspace{5mm}
       ${}^2${\it Division for Algebra and Geometry, Department of
         Mathematical Sciences,\\
         Gothenburg University and Chalmers University of Technology,\\
 SE-412 96 Gothenburg, Sweden}

\end{center}

\vfill

\begin{quote}

\textbf{Abstract:} Tensor hierarchy algebras constitute a class of
non-contragredient Lie superalgebras, whose finite-dimensional
members are the ``Cartan-type'' Lie superalgebras in Kac's classification.
They have applications in mathematical physics, especially in extended
geometry and gauged supergravity. We 
further develop the recently proposed
definition of tensor hierarchy algebras
in terms of generators and
relations encoded in a Dynkin diagram (which coincides with the
diagram for a related Borcherds superalgebra). 
We apply it
to cases where a grey node is added to the Dynkin diagram
of a rank $r+1$ Kac--Moody algebra $\fg^+$, which in turn 
is an extension of a
rank $r$ finite-dimensional semisimple simply laced Lie algebra
$\fg$.
The algebras are specified by $\fg$ together with a dominant
integral weight $\lambda$.
As a by-product, a remarkable identity involving representation
matrices for arbitrary integral highest weight representations of
$\fg$ is proven.
An accompanying paper \cite{CederwallPalmkvistTHAII}
describes the application of tensor hierarchy algebras
to the gauge structure and dynamics in models of extended geometry.

\end{quote} 
\vfill

\hrule

\noindent{\tiny email:
  martin.cederwall@chalmers.se, jakob.palmkvist@chalmers.se}

\newpage

\tableofcontents

\pagestyle{plain}

\section{Introduction}

Simple finite-dimensional Lie superalgebras were classified by Kac in ref.
\cite{Kac77B}.
Among them are the peculiar Cartan-type
superalgebras $W(n)$ and $S(n)$, where $W(n)$ is
the derivation algebra of the associative superalgebra of (point-wise) forms
in $n$ dimensions under the wedge product (the Grassmann algebra on
$n$ generators), and $S(n)\subset W(n)$ is a scale-preserving
subalgebra.
These superalgebras are non-contragredient, meaning that they do not
have a presentation in terms of
generators and relations which is symmetric, up to signs, under the interchange of generators at positive and negative levels.

In ref. \cite{Carbone:2018xqq}, we introduced a set of generators and
relations for $W(n)$ and $S(n)$, with an antisymmetry
between positive and negative levels, by modifying the presentation of the contragredient Lie superalgebra
$A(0,n-1)=\mathfrak{sl}(1|n)$.
This construction starts with the Dynkin diagram of $A(0,n-1)=\mathfrak{sl}(1|n)$
but can be applied to other similar Dynkin diagram as well. In general it leads to a so called 
tensor hierarchy algebra (THA) \cite{Palmkvist:2013vya}, a Lie superalgebra that is an infinite-dimensional
super-extension of a Kac--Moody algebra $\fg$.
The Cartan-type superalgebras of Kac are obtained as the
special cases $W(A_{n-1})=W(n)$ and $S(A_{n-1})=S(n)$. The corresponding contragredient Lie superalgebra is
a Borcherds superalgebra $\scr B(\fg)$ such that $\scr B(A_{n-1})=A(0,n-1)$.
In ref. \cite{Carbone:2018xqq} we studied in detail the case of finite-dimensional $\fg$. 
The main purpose of the present paper is to extend this study to the case where $\fg$ is extended by an additional node in the Dynkin diagram
to a possibly infinite-dimensional Kac--Moody algebra $\fg^+$
(the
precise definition depends in addition to $\fg$ on the choice of a
dominant integral weight $\lambda$, in a way that will be clarified later).

The invention of the THA's was motivated by the need to accommodate the
embedding tensor of gauged supergravities in the algebra \cite{Palmkvist:2013vya,Greitz:2013pua}. It has
subsequently become clear \cite{Bossard:2017wxl,Bossard:2017aae,Cederwall:2018aab,Bossard:2019ksx} that they are also needed as an algebraic basis
for models of extended geometry \cite{Cederwall:2017fjm}.
In certain
simple cases, where so called ancillary transformations do not appear,
only the corresponding Borcherds superalgebra is needed. In 
ref. \cite{Cederwall:2018aab} we derived an $L_\infty$-algebra
from it, encoding the gauge structure in the absence of ancillary transformations.
The more general situation demands that a
THA is used. We refer to the accompanying paper
\cite{CederwallPalmkvistTHAII} for details on extended geometry, and
for details about gauge transformations (generalised diffeomorphisms)
and dynamics in such models.

The paper is organised as follows. In Section \ref{BorcherdsSection},
we review the Chevalley--Serre construction of the corresponding Borcherds
superalgebras $\BB(\fg^+)$. This presentation is then generalised,
using the same Dynkin diagram, to the THA's in Section
\ref{THASection}. Section \ref{TensorProductSection} deals with the
tensor product between the adjoint of $\fg$ and any highest weight
representation, using the multiplicity formula of
Parthasarathy, Ranga Rao and Varadarajan
\cite{ParthasarathyEtAl}. This tensor product is needed to determine
the content of a THA in a double grading, where each grade
forms a $\fg$-module. A $\fg$-covariant description is then given in
Section \ref{gReprSection}, and  
a sequence of subalgebra embeddings
of THA's is described in Section \ref{EmbeddingSection}.
The $\fg$-covariant description leads to a remarkable algebraic
identity involving projectors on irreducible submodules of the tensor
product $R(\lambda)\otimes\adj$,
which is verified explicitly in a series of examples in
Section \ref{ExamplesSection}. We end with conclusions in Section \ref{consec}.

The accompanying paper \cite{CederwallPalmkvistTHAII} deals with the
application of the tensor hierarchy algebras $S(\fg^+)$ constructed here to
extended geometry, both the gauge structure (in the form of an
$L_\infty$ algebra) and the dynamics. 
In order for both
papers to be reasonably self-contained, their contents have a certain overlap. 

\section{The Borcherds superalgebra $\BB$\label{BorcherdsSection}}

We start with a finite-dimensional semisimple Lie algebra $\fg$ or rank $r$, which we assume to be simply laced, and a dominant integral weight $\lambda$,
which we assume satisfies $(\lambda,\lambda)\neq 1$ in a normalisation
where the simple roots $\alpha_i$ of $\fg$ have length squared
$(\alpha_i,\alpha_i)=2$. 
The assumption that $\lambda$ is dominant integral means that the Dynkin labels $\lambda_i =(\lambda,\alpha_i)$ are non-negative integers (not all zero).
The Dynkin labels are the coefficients of $\lambda$ in the basis of
fundamental weights $\Lambda_i$, defined by
$(\Lambda_i,\alpha_j)=\delta_{ij}$.
A dominant integral weight $\lambda$ defines a highest weight representation,
which is denoted $R(\lambda)$, with $\lambda$ as highest weight, The
dual (conjugate) representation with lowest weight $-\lambda$ is
denoted $R(-\lambda)=\overline{R(\lambda)}$.
We use the same notation for the representations and the corresponding
modules. In concrete examples they may also be denoted by their
dimension, written in boldface.

\begin{figure}
  \centerline{\includegraphics{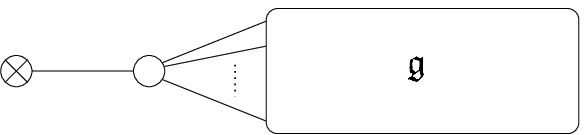}\hskip5mm\includegraphics{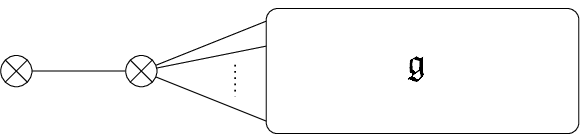}}
  \caption{\it Two equivalent Dynkin diagrams for $\BB(\fgrplus)$, $W(\fgrplus)$
    and $S(\fgrplus)$. Removing the ``grey'' node in the first diagram
    yields the Dynkin diagram of $\fg^+$.} 
  \label{DynkinFigure}
\end{figure}

The Borcherds superalgebra $\scr B=\scr B(\fg^+)$ can be constructed
by adding two nodes to the Dynkin diagram of $\fg$. This can be done
in two different but equivalent ways, related by an ``odd Weyl reflection''
\cite{Dobrev:1985qz} as shown in Figure \ref{DynkinFigure}.
In ref. \cite{Cederwall:2018aab} we considered $\scr B$ as constructed
from a Dynkin diagram of the second type, with two grey nodes. 
Here we will instead construct $\scr B$ from a Dynkin diagram of the
first type, with only one grey node. A difference in notation compared to
ref. \cite{Cederwall:2018aab} is that we label the $r$ nodes in the
Dynkin diagram of $\fg$ 
(or the corresponding simple roots) by an index $i$ that takes the
values $i=2,\ldots,r+1$ rather than $i=1,2,\ldots,r$. 

Thus, to the Dynkin diagram of $\fg$ we first 
add a white node $1$ connected to 
node $i$
by $\lambda_i$ lines, extending $\fg$ to $\fg^+$. This first
extending node corresponds to a simple root $\alpha_1$ of the same length as the
simple roots of $\fg$ (even when it is connected with multiple lines, which means that there are no arrows).
Then, the Dynkin diagram of $\fg^+$ is extended by a grey node $0$
connected to node $1$ with a single line, and disconnected from all
nodes $i$ 
of $\fg$.
The corresponding simple root $\alpha_0$ is a null root.
This Dynkin diagram corresponds to a Cartan matrix $B_{ab}$ ($a,b=0,1,\ldots,r+1$) where $B_{ij}$ ($i,j=2,3,\ldots,r+1$)
is the Cartan matrix of $\fg$ and
\begin{align}
B_{1i}=B_{i1}&=-\lambda_i\;, &
B_{01}&=B_{10}=-1\;,& B_{11}&=2\;, & B_{0i}=B_{i0}&=B_{00}=0\;.
\end{align}

To each node $a$ we associate three generators $e_a,f_a,h_a$. Among these $3(r+2)$ generators, $e_0$ and $f_0$ are odd, the others even.
Now $\scr B$ is defined as the Lie superalgebra generated by the set $\{e_a,f_a,h_a\}$ modulo the Chevalley--Serre relations
\begin{align}
[h_a,e_b]=B_{ab}e_b\;,  \qquad   
[h_a,f_b]=-B_{ab}f_b\;, \qquad   
[e_a,f_b]=\delta_{ab}h_b\;,  \label{eigen0}
\end{align}
\begin{align}
({\rm ad}\,e_a)^{1-B_{ab}}(e_b)=({\rm ad}\,f_a)^{1-B_{ab}}(f_b)=0\;.
\label{serre0}
\end{align}
Note that we use the notation $[\cdot,\cdot]$ for the brackets, also
between two odd (fermionic) elements, when it is symmetric.

When we extend $\fg$ to $\scr B$ we also extend the Cartan subalgebra $\mathfrak{h}$ of $\fg$ to a Cartan subalgebra $\scr H$ of $\scr B$.
The set of simple roots $\alpha_a$ constitute a basis of the dual space $\scr H^\ast$ with an inner product given by the Cartan matrix,
$(\alpha_a,\alpha_b)=B_{ab}$.
Since we assume that $\fg$ is simply laced, the Cartan matrix $B_{ij}$ is symmetric and
all the simple roots have the same length squared, which we normalise to $2$. It should however
be straightforward to generalise our results to situations where $\fg$
is not simply laced, as long as $\lambda$ has vanishing Dynkin labels for the short roots (\ie, if node $1$ is disconnected
from nodes representing short roots). We write $\psi(\alpha)=h_\alpha$
for the isomorphism $\psi: \scr {\mathfrak{h}}^\ast \to \mathfrak{h}$
given by $\psi(\alpha_i)=h_i$. 

The Borcherds superalgebra $\scr B$ can be decomposed into a direct sum of subspaces, labelled by pairs of integers $(p,q)$
where $p$ and $q$ are the eigenvalues of $-h_0$ and
\begin{align}
\big((\lambda,\lambda)-1\big)h_0-h_1-h_\lambda\;,
\end{align}
respectively. We will refer to them as {\it level} and {\it height}, respectively.\footnote{We will occasionally talk about ``levels'' 
with respect to other $\mathbb{Z}$-gradings too, 
and also about the ``height'' of a root or a weight in the usual meaning.}
This is a consistent $(\mathbb{Z}\times \mathbb{Z})$-grading in the sense that the subspaces at even and odd $p+q$ belong
to the even and odd part of the Lie superalgebra, respectively.
Since $\fg$ is a subalgebra at $(p,q)=(0,0)$,
the subspace at any definite pair of integers $(p,q)$
forms an $\fg$-module.
Our notation for these modules is given in Table \ref{GeneralTable}. As can be seen there, all modules come in pairs,
except for those at level $p=0$.
For all other pairs of integers, any irreducible module that appears at $(p,q)$ also appears at either $(p,q+1)$ or $(p,q-1)$.
This ``doublet structure'' follows from
the fact that $e_0,f_0,h_0$ form a Heisenberg superalgebra,
\begin{align}
[e_0,f_0]=[f_0,e_0]&=h_0\;, & [h_0,e_0]=[h_0,f_0]&=0\;,
\end{align}
that commutes with $\fg$.
In ref. \cite{Cederwall:2018aab} we defined corresponding raising and
lowering operators. In the notation that we use here, the definitions 
take the form\footnote{Note that, unlike in
  ref. \cite{Cederwall:2018aab}, the raising operator is here
  associated with an ``$f$ generator'', and the lowering operator with
  an ``$e$ generator''. 
This is a consequence of the ``odd Weyl reflection'' that relates the
two diagrams in Figure \ref{DynkinFigure} to each other.} 
\begin{align}
\sh \ :\quad A &\mapsto A^\sh = -\frac1p [A,f_0]\;,\nn\\
\fl \ :\quad  A &\mapsto A^\fl = -[A,e_0]\;, \label{hojningsankning}
\end{align}
for any element $A$ at level $p\neq0$.
It follows from the Chevalley--Serre relations that they satisfy
\begin{align}
\sh^2 &= \fl^2 = 0\;, & \sh\fl+\fl\sh&=\id\;,
\label{hojningsankningegenskaper}
\end{align}
and commute with the adjoint action of any element in $\fg$.
We introduce basis elements $E_M$ and $F^M$ for the odd subspaces at
$(p,q)=(1,0)$ and $(p,q)=(-1,0)$, respectively, which form the
$\fg$-modules $R(-\lambda)$ and $R(\lambda)$.
Accordingly, $E_M^\sh$ and $F^{\fl M}$ (denoted $\wt E_M$ and $\wt
F^M$ in refs.
\cite{Palmkvist:2015dea,Cederwall:2018aab}) are basis elements for 
the even subspaces at $(p,q)=(1,1)$ and $(p,q)=(-1,-1)$. For the subalgebra $\fg$ at $(p,q)=(0,0)$ we introduce basis elements $T_\alpha$,
where the adjoint index can be raised by the inverse $\eta^{\alpha\beta}$ of the Killing form $\eta_{\alpha\beta}=(T_\alpha,T_\beta)$.
At $(p,q)=(0,0)$ we also have a two-dimensional abelian subalgebra that commutes with $\fg$. As basis elements, it is convenient to choose
$k=h_0+h_1+h_\lambda$ and $\wt k=h_1+h_\lambda$. Also the generators $e_0$ and $f_0$ at $(p,q)=(0,-1)$ and $(p,q)=(0,1)$, respectively,
are clearly singlets under $\fg$ since node $0$ is disconnected from the nodes $2,3,\ldots,r+1$. At levels $p=0,\pm1$ (the {\it local part} of the Lie superalgebra with respect to this $\mathbb{Z}$-grading)
we thus have the basis elements shown in Table
\ref{GeneralBTableBasis}.
\begin{table}[h]
\begin{align*}
\xymatrix@!0@C=2.2cm{
\cdots \ar@{-}[]+<1.1cm,1em>;[dddddd]+<1.1cm,-1em> \ar@{-}[]+<-0.8cm,-1em>;[rrrrrr]+<0.6cm,-1em>&
p=-1 \ar@{-}[]+<1.1cm,1em>;[dddddd]+<1.1cm,-1em>&
p=0 \ar@{-}[]+<1.1cm,1em>;[dddddd]+<1.1cm,-1em>&
p=1\ar@{-}[]+<1.1cm,1em>;[dddddd]+<1.1cm,-1em>&
p=2\ar@{-}[]+<1.1cm,1em>;[dddddd]+<1.1cm,-1em>&
p=3\ar@{-}[]+<1.1cm,1em>;[dddddd]+<1.1cm,-1em>&\cdots\\ \cdots&&&&&& *+[F-:red][red]{ n=0}\\
q=3 &&          &           &                    &     {{\ttR}}_3 \ar@{-}@[red][ur]& *+[F-:red][red]{ n=1}\\
q=2&&          &           &        {\tR_2} \ar@{-}@[red][ur]&        {\tR}_3 \oplus {{\ttR}}_3 \ar@{-}@[red][ur] & *+[F-:red][red]{ n=2 } \ar@{-}@[red][dl]\\
q=1&&{\bf 1} \ar@{-}@[red][dl]  &     R_1 \ar@{-}@[red][ur] &   R_2
\oplus \tR_2 \ar@{-}@[red][ur] &   {R}_3 \oplus {{\tR}}_3
& *+[F-:red][red]{ n=3 } \ar@{-}@[red][dl]\\
q=0&\overline R_1 &{\bf 1}\oplus{\bf adj}\oplus{\bf 1}
\ar@{-}@[red][ur]  \ar@{-}@[red][dl]  & R_1
\ar@{-}@[red][dl] \ar@{-}@[red][ur] & R_2
\ar@{-}@[red][ur]&  R_3
&\cdots\\
\cdots\ar@{-}@[red][ur] &\overline R_1 & {\bf 1} & &
&
&
}
\end{align*}
\caption{\it The general structure of the superalgebra
  ${\scr B}(\fg^+)$.
  Red lines are the usual levels $n=p-q$ in the level
  decomposition of ${\scr B}(\fg^+)$, and form $\fg^+$-modules.}
\label{GeneralTable}
\end{table}
\begin{table}
  \begin{align*}
  \xymatrix@=.4cm{
    \ar@{-}[]+<2.2em,1em>;[ddd]+<2.2em,-1em>
    \ar@{-}[]+<-0.8cm,-1em>;[rrr]+<1.4cm,-1em>
    &\ar@{-}[]+<2.3em,1em>;[ddd]+<2.3em,-1em> p=-1
    & \ar@{-}[]+<2.9em,1em>;[ddd]+<2.9em,-1em> p=0 &p=1\\
q=1&&f_0& E_M^\sh
       \\ 
q=0&F^M
&k\quad{T_\alpha}\quad\tk
           & E_M\\
q=-1 & F^{\fl M}& e_0 
  }
\end{align*}
  \caption{\it Basis elements for $\sB(\fgrplus)$ at $p=-1,0,1$.}
\label{GeneralBTableBasis}
\end{table}

Of particular interest are modules $R_2$ and $\tR_2$. $R_2$ contains
the symmetric tensor product of two $R_1$'s, except the lowest one, which is
removed by the relation $[e_0,e_0]=0$, so that
\begin{align}
R_2=\vee^2R(-\lambda)\ominus R(-2\lambda)\;.
\end{align}
$\tR_2$ contains the antisymmetric tensor product of two $R(-\lambda)$'s,
with the modules corresponding to Serre relations in $\fg^+$
containing two $e_1$'s removed, \ie,
\begin{align}
  \tR_2=\wedge^2R(-\lambda)
  \ominus\bigoplus\limits_{i:\lambda_i=1}R(-(2\lambda-\alpha_i))
\end{align}
(we use $\vee$ and $\wedge$ for symmetric and antisymmetric tensor products).

The (anti-)commutation relations are
\begin{align}
[T_\alpha,E_{M}]&=-(t_\alpha)_{M}{}^{N} E_{N}\;, &
[T_\alpha,E_{M}^\sh]&=-(t_\alpha)_{M}{}^{N}E_{N}^\sh\;,\nn\\ 
[k,E_{M}]&=-(\lambda,\lambda)E_{M}\;,  &
[\tk,E_{M}^\sh]&=(2-(\lambda,\lambda))E_{M}^\sh\;,\nn\\ 
[\tk,E_{N}]&=(1-(\lambda,\lambda))E_{N}\;, &
[k,E_{N}^\sh]&=(1-(\lambda,\lambda))E_{N}^\sh\;,\nn\\ 
[f_0,E_{N}]&=-E_{N}^\sh\;, & [f_0,E_{N}^\sh]&=0\;,\nn\\
[e_0,E_{N}]&=0\;, & [e_0,E_{N}^\sh]&= E_{N}\;,
\label{borcherds-comm-rel}
\end{align}
\begin{align}
[T_\alpha,F^{N}]&=(t_\alpha)_{M}{}^{N} F^{M}\;, &
[T_\alpha,F^{\fl N}]&=(t_\alpha)_{M}{}^{N}F^{\fl M}\;,\nn\\ 
[k,F^{N}]&=(\lambda,\lambda)F^{N}\;, &
[\tk,F^{\fl N}]&=((\lambda,\lambda)-2)F^{\fl N}\;,\nn\\ 
[\tk,F^{N}]&=((\lambda,\lambda)-1)F^{N}\;, &
[k,F^{\fl N}]&=((\lambda,\lambda)-1)F^{\fl N}\;,\nn\\ 
[f_0,F^{N}]&=0\;, & [f_0,F^{\fl N}]&=-F^N\;,\nn\\
[e_0,F^{N}]&=-F^{\fl N}\;, & [e_0,F^{\fl N}]&= 0\;,\label{borrel2}
\end{align}
\begin{align}
[E_{M},F^{N}]&=-(t^\alpha)_{M}{}^{N} T_\alpha + \de_{M}{}^{N} k\;, &
[E_{M}^\sh,F^{\fl N}]&=-(t^\alpha)_{M}{}^{N} T_\alpha + \de_{M}{}^{N}
\tk\;, 
\nn\\
[E_{M},F^{\fl N}]&=\de_{M}{}^{N} e_0\;, & [E_{M}^\sh, F^{N}]&=\de_{M}{}^{N} f_0\;. \label{borrel3}
\end{align}

\section{Modifying $\scr B$
to a tensor hierarchy algebra\label{THASection}}
In the Borcherds superalgebra, there is never a nontrivial module
$\tR_1$. A direct motivation from extended geometry to introduce a
tensor hierarchy algebra comes from the need for such a module in
order to describe ancillary transformations \cite{CederwallPalmkvistTHAII}. 

In ref. \cite{Carbone:2018xqq}, two different Lie superalgebras
$W(\fg^+)$ and $S(\fg^+)$, 
both called tensor hierarchy algebras, were defined in the case of finite-dimensional $\fg^+$. We will here give a slightly different definition, valid also for infinite-dimensional
$\fg^+$ (but still finite-dimensional $\fg$).
The algebra needed in extended geometry
\cite{CederwallPalmkvistTHAII} is $S(\fg^+)$, but in accordance with
ref. \cite{Carbone:2018xqq} we first give the definition of $W(\fg^{+})$,
and then explain how the definition of $S(\fg^{+})$ is obtained from it. 

Our investigation will exclude the case $(\lambda,\lambda)=1$, which
happens when $\fg=D_r$ and $\lambda=\Lambda_1$, so that $R(\lambda)$
is the vector representation.
This case is somewhat degenerate (see eqs. \eqref{HEBrackets}
and \eqref{MuAlphaEq}),
for example in the sense that the
embeddings of Section \ref{EmbeddingSection} are not valid.
The corresponding tensor hierarchy algebras are still well-defined,
and should be relevant for double geometry. However, some aspects, especially
the identification of ideals, require a
special treatment, which we will not deal with here.

\subsection{The tensor hierarchy algebra $W$\label{WSubSec}}

The tensor hierarchy algebras $W=W(\fg^+)$ and $S=S(\fg^+)$ are defined from the same
Dynkin diagram and Cartan matrix $B_{ab}$ as $\scr B$, corresponding to an $(r+2)$-dimensional
vector space with a basis consisting of simple roots $\alpha_a$ and inner product $(\alpha_a,\alpha_b)=B_{ab}$.
However, the assignments of generators to the nodes in the Dynkin diagram is different.

The generators of $W$ are obtained from those of $\scr B$ in the following way.
The even generators $e_i$, $f_i$, $h_a$ and the odd generator $e_0$ are kept, but the other odd generator $f_0$ 
is replaced by $r+1$ odd generators $f_{0a}$, where $a=0,2,\ldots,r+1$.
Henceforth, whenever $f_{0a}$ appears we assume $a\neq 1$, and whenever $f_{a}$ appears we assume
$a\neq0$. Otherwise, if nothing else explicitly stated, the indices $a,b,\ldots$ will take the values $0,1,2,\ldots,r+1$.
The default values of the indices $i,j,\ldots$ will be $2,3,\ldots,r+1$.
We introduce a consistent $(\mathbb{Z} \times \mathbb{Z})$-grading with level $p$ and height $q$ as for $\scr B$.

In the definition of $W$ we now first define an auxiliary algebra $\wt W$ as the Lie superalgebra generated by
the set $\{e_a,f_a,f_{0a},h_a\}$
modulo the relations
\begin{align}
[h_a,e_b]=B_{ab}e_b\;,  \qquad   
[h_a,f_b]=-B_{ab}f_b\;, \qquad   
[e_a,f_b]=\delta_{ab}h_b\;,  \label{eigen1}
\end{align}
\begin{align}
({\rm ad}\,e_a)^{1-B_{ab}}(e_b)=({\rm ad}\,f_a)^{1-B_{ab}}(f_b)=0\;.
\label{serre1}
\end{align} 
\begin{align}
[e_0,f_{0a}]&=h_a\;, & 
[h_a,f_{0b}]&=-B_{a0}f_{0b}\;, &
[e_i,[f_j,f_{0a}]]&= \delta_{ij}B_{aj}f_{0j}\;, \label{eifjf0a}
\end{align}
\begin{align}
[e_1,f_{0a}]=
[f_1,[f_1,f_{0a}]]=
[f_{0a},f_{0b}]&=0\;.
\label{IdealJ2'}
\end{align}
In the first two lines
we recognise the relations (\ref{eigen0}) (but now with the assumption that the single index on $f$ does not take the value $0$).

Let $\wt W_{(i,j)}$ be the subspace of $\wt W$ spanned by all elements of the form
\begin{align}
[x_1,[x_2,\,\ldots,[x_{N-1},x_N]\cdots]] \label{multibracket}
\end{align}
for some integer $N$, where each  $x_j \in \{e_a,f_a,f_{0a},h_a\}$ ($j=1,2,\ldots,N)$
and 
among the $N$ elements $x_j$, the generators $e_1$ and $f_1$ appear $i$ and $j$, times, respectively.
(Henceforth, we will occasionally write a multi-bracket of the form (\ref{multibracket}) simply as
$[x_1,\,\ldots,x_{N-1},x_N]$.) The algebra $\wt W$ has a $\mathbb{Z}$-grading $\wt W = \bigoplus_{p \in \mathbb{Z}} \wt W_p$ where 
$\wt W_p$ is the sum of all subspaces $\wt W_{(i,j)}$ such that $i-j=p$.
Let $J$ be the maximal ideal of $\wt W$ intersecting
$\wt W_0$
trivially (obtained by taking the sum of all ideals with this property). We define $W$ as the quotient obtained from $\wt W$ by factoring out this ideal, $W=\wt W/J$.

We will see that $\wt W_{(1,0)}=\wt W_1$ and $\wt W_{(0,1)}=\wt W_{-1}$. This is not obvious. Since there are no relations $[e_i,f_{0a}]=0$ for $i=2,3,\ldots,r+1$,
the Lie superalgebra $\wt W$ does not admit a triangular decomposition. 
When we consider basis elements of the form (\ref{multibracket}) for $N\geq 2$, we cannot assume 
that either all $x_j \in \{e_a\}$ or all $x_j \in \{f_a,f_{0b}\}$. Moreover, if one of the elements $x_j$ is equal to $e_a$ and another one is equal to
$f_{a}$ (if $a \neq 0$) or some $f_{0b}$ (if $a=0$), then it is in general not possible to rewrite any such expression using $[e_a,f_b]=\delta_{ab}f_b$
or $[e_0,f_{0b}]=h_b$
so that both disappear. It is however possible in special cases:
for any $a$ when $\fg$ is finite-dimensional \cite{Carbone:2018xqq}
and, as we will see, for $a=1$ when $\fg^+$ is
finite-dimensional. (What we will show explicitly is the corresponding
statement for the subalgebra $S$, but it can be shown in the same way
for $W$.) 

In ref. \cite{Carbone:2018xqq}, where $\fg$ was assumed to be
finite-dimensional and $\lambda$ a fundamental weight
$\lambda=\Lambda_k$, the tensor hierarchy algebra $W$ was defined
similarly from an auxiliary algebra $\wt W$, but with the
$\mathbb{Z}$-grading associated to node $0$ rather than to node
$1$. It was then shown that,  
in the case of $\fg=A_r$ and $\lambda=\Lambda_2$,
where $W$ is the finite-dimensional Lie superalgebra of Cartan type $W(r+2)$, the ideal $J$ intersecting the local part trivially was generated by the relations
\begin{align}
[f_{0a},f_{0b}]=[f_{0i},[f_{0j},f_1]]=[(f_{02}-f_{00}),[f_{0j},f_1]]=0 \label{gamlaserre}
\end{align}
for $i,j=3,\ldots,r+1$.
Here we have instead included the relations $[f_{0a},f_{0b}]=0$ 
already in the definition of $\wt W$ and the ideal that we factor out is the maximal one intersecting $\wt W_0$ trivially,
where the $\mathbb{Z}$-grading is associated to node $1$ rather than node $0$ (the relations involving $f_1$ are contained in this ideal). The reason is that we have $\wt W_{(1,0)}=\wt W_1$ and $\wt W_{(0,1)}=\wt W_{-1}$ in this
$\mathbb{Z}$-grading, as discussed above.

Another difference in comparison with the relations in
ref. \cite{Carbone:2018xqq} is that, among the relations 
\begin{align}
[e_a,[e_a,f_{0b}]]=
[f_a,[f_a,f_{0a}]]&=0\;,
\label{dubbel}
\end{align}
there, we have only included
$[f_1,[f_1,f_{0a}]]=0$ here. The other ones follow in fact from the relations above.
For $[e_a,[e_a,f_{0b}]]=0$ with $a=0$ or $a=1$, this was noted already
in ref. \cite{Carbone:2018xqq},
and also that 
$[e_i,f_{0a}]=[f_i,f_{0a}]=0$ if $B_{ia}=0$. 
Suppose now that $B_{ij}=-1$. Then
\begin{align}
2[e_i,[e_i,f_{0j}]]&=[e_i,[e_i,[e_j,[f_j,f_{0j}]]]]\nn\\
&=2[e_i,[e_j,[e_i,[f_j,f_{0j}]]]]-[e_j,[e_i,[e_i,[f_j,f_{0j}]]]]=0\;
\end{align}
and finally
\begin{align}
-[e_i,[e_i,f_{0i}]]&=[e_i,[e_i,[e_i,[f_i,f_{0j}]]]]\nn\\
&=[e_i,[e_i,[f_i,[e_i,f_{0j}]]]]\nn\\
&=[e_i,[h_i,[e_i,f_{0j}]]]+[e_i,[f_i,[e_i,[e_i,f_{0j}]]]]=0\;.
\end{align}
In the same way, one can show that $[f_i,[f_i,f_{0a}]]=0$.

\subsection{The tensor hierarchy algebra $S$}

It is easy to see that
if we remove the generators $f_{0i}$ and the relations that involve them,
but keep $f_{00}$, then we recover $\scr B$ from $W$ (identifying $f_{00}$ with $f_0$).
Conversely, we can remove the generators $f_{00}$ and $h_0$ and the relations that involve them, but keep $f_{0i}$. Then we obtain the tensor hierarchy algebra
$S$. Thus $S$ is defined, via an auxiliary algebra $\wt S$, in the same way as $W$ above but without the generators $f_{00}$ and $h_0$ and the relations that involve them. We assign values of $p$ and $q$ to the generators as in $W$.

In $W$ we can define operators $\sh$ and $\fl$ satisfying
(\ref{hojningsankningegenskaper}), 
by replacing $f_0$ by $f_{00}$ in (\ref{hojningsankning}). In $S$ this
is not possible 
since there is no generator $f_{00}$ in $S$ that could be identified
with $f_0$ in $\scr B$.  
One might think that this would mean an absence in $S$ of the ``doublet
structure'' present in $\scr B$ at nonzero levels. However, it is in fact still present in 
$S$ (we do not have a proof to all levels in the general
case, but the opposite seems extremely unlikely), and it even extends to level $p=0$. 

We will show that it is indeed possible to define an operator $\sh$ on the subalgebra of $S$
generated by $\{e_i,f_i,e_0,f_{0i},h_a\}$
such that $\sh$ satisfies (\ref{hojningsankningegenskaper}), with $\fl$
still defined by (\ref{hojningsankning}). First we set
\begin{align}
h_i{}^\sh&=-f_{0i}\;,& h_1{}^\sh&=f_{0\lambda}\;, & e_0{}^\sh=\wt k&=h_1+h_\lambda\;, & f_{0i}{}^\sh&=0\;.
\end{align}
It then follows that $\sh^2=0$ and $\sh\fl+\fl\sh=\id$ on these generators, and that $[x,e_0]^\sh=[x,e_0{}^\sh]$,
where $x$ is any element in $\fg$.

Let us write $h_\alpha{}^\sh=-f_{0\alpha}$.
In order to extend the operator $\sh$ to the root vectors $e_\alpha$ of $\fg$ (corresponding to positive or negative roots), we note that
\begin{align}
(\alpha,\beta)[e_\alpha,f_{0\gamma}]=(\alpha,\gamma)[e_\alpha,f_{0\beta}]\; \label{fundamentalrelation}
\end{align}
for any root $\alpha$ of $\fg$ and $\beta,\gamma \in \mathfrak{h}^\ast$. This
was shown in ref. \cite{Carbone:2018xqq} in the case when $\alpha,\beta,\gamma$ are simple roots, and it is straightforward to show it also in this general case.
We can then unambiguously set
\begin{align}
e_\alpha{}^\sh &= \frac1{(\alpha,\beta)}[e_\alpha,f_{0\beta}]
\end{align}
for any root $\alpha$ 
of $\fg$ and any $\beta \in \scr \fg^\ast$ such that
$(\alpha,\beta)\neq0$. As shown in ref. \cite{Carbone:2018xqq} (with $\beta=\varrho$), this implies that 
\begin{align}
[x,y^\sh]&=[x,y]^\sh \label{adjoint}
\end{align}
for any $x,y \in \fg$. We also set $e_1{}^\sh =0$ and
\begin{align}
f_1{}^\sh &= \frac1{(\lambda,\lambda)}[f_1,f_{0\lambda}]\;.
\end{align}
Another result from ref. \cite{Carbone:2018xqq} that we will use is
\begin{align}
[e_{\alpha},[e_{-\alpha},f_{0\beta}]]=(\alpha,\beta)f_{0\alpha}  \label{fundamentalrelation2}
\end{align}
for any root $\alpha$ of $\fg$ and any $\beta\in \mathfrak{h}^\ast$.

\subsection{Local part of $S$}

We will now study the subspaces of $S$ at levels $p=0,\pm1$ and
decompose each of them further into subspaces at different heights
$q$. 
It will be useful to consider also a $\mathbb{Z}$-grading of $\fg$
with respect to $\lambda$. 
We let $\fg_{(\ell)}$ be the subspace of $\fg$ spanned by all root
vectors $e_\alpha$ corresponding  
to roots $\alpha$ such that $(\alpha,\lambda)=\ell$, and, if $\ell=0$,
the Cartan generators 
$h_i$ of $\fg$. 
We thus have $\fg=\bigoplus_{\ell \in \mathbb{Z}} \fg_{(\ell)}$.
We also write, for example, $\fg_{(\leq1)}=\bigoplus_{\ell\leq 1}\fg_{(\ell)}$. For homogeneous elements $x$ in $\fg$ with respect to this $\mathbb{Z}$-grading
we call this degree $\lambda$-level and denote it by $\ell(x)$, so that $x \in \fg_{(\ell(x))}$. The Dynkin diagram of $\fg_{(0)} \subseteq \fg$ is obtained by
removing the nodes $i$ in the Dynkin diagram of $\fg$ that are
connected to node $1$ in the extension to $\fg^+$, \ie, the nodes with
$\lambda_i\neq0$. 

In the notation introduced above for $\wt W$, the algebra $\wt S$ contains subspaces $\wt S_{(0,0)}$, $\wt S_{(1,0)}$ and $\wt S_{(0,1)}$
at levels $0$, $1$ and $-1$, respectively. The subspace $S_{(0,0)}$ is the subalgebra generated by 
all generators but $e_1$ and $f_1$. We will also denote it by $S'$ below. The subspace $S_{(1,0)}$ is spanned by multi-brackets that contain precisely
one $e_1$ and no $f_1$, whereas, conversely, $S_{(0,1)}$ is spanned by multi-brackets that contain precisely
one $f_1$ and no $e_1$. In the multi-brackets that span $S_{(1,0)}$, the only $e_1$ generator can always be put in the innermost position by the Jacobi identity.
When considered as spanned by such multi-brackets, we say that the $S_{(1,0)}$ is the $S'$-module generated by $e_1$, and denote it by $S'(e_1)$. We will use
the corresponding notation, $\fg(a)$, for the $\fg$-module generated
by some element $a$ in $S$ (or in the algebra currently under investigation).

At this point it is not clear that the algebras $\wt W$ and $\wt S$ are non-trivial, \ie\ that the 
relations (\ref{eigen1})--(\ref{IdealJ2'}) generate a proper ideal of the free Lie superalgebra generated 
by $\{e_a,f_a,f_{0a},h_a\}$ and not the whole free Lie superalgebra itself. This will be shown in Section \ref{gReprSection}.
We will anticipate this result and proceed under the assumption that $\wt W$ and $\wt S$ indeed are non-trivial.

\subsubsection{The subalgebra $S'$}
We start by examining the contents of the subalgebra $S'$ of $S$.
At height $q=0$, it contains the subalgebra generated by $\{e_i,f_i,h_1\}$. 
This is 
$\fg \oplus \langle \wt k \rangle$ of $\fg$, the direct sum of $\fg$ and a one-dimensional Lie algebra spanned by $\wt k$.
At height $q=1$ and $q=-1$ it contains the $\fg$-modules $\fg(e_0)$ and $\fg(\mathfrak{h}^\sh)$  generated by $e_0$ and all $f_{0i}$, respectively. The first one is a singlet
since $e_0$ commutes with all $e_i,f_i$. The second one is $\fg^\sh$, which is isomorphic to $\fg$ itself, the adjoint module, according to (\ref{adjoint}).
Since $[e_0,\fg^\sh]=\fg$, there is no other $\fg$-module in $S'$ at height $q=0$ or $q=\pm1$. 
Furthermore, since $[e_0,e_0]=0$, the algebra $S'$ does not contain any non-trivial element at height $q\leq -2$. 
To see that $S'$ does not contain any non-trivial element at height $q\geq 2$ either, we use (\ref{fundamentalrelation}). We then get
$[[e_\alpha,f_{0\beta}],f_{0\gamma}]=[e_\alpha,[f_{0\beta},f_{0\gamma}]]=0$ if $(\alpha,\gamma)=0$, and otherwise
\begin{align}
[[e_\alpha,f_{0\beta}],f_{0\gamma}]=\frac{(\alpha,\beta)}{(\alpha,\gamma)}[[e_\alpha,f_{0\gamma}],f_{0\gamma}]=\frac{(\alpha,\beta)}{2(\alpha,\gamma)}[e_\alpha,[f_{0\gamma},f_{0\gamma}]]=0\;.
\end{align} 
From this it easily follows that $[\fg^\sh,\fg^\sh]=0$. We summarise:
\begin{align} \label{S'}
  S' &= \langle e_0 \rangle \oplus \langle \wt k \rangle \oplus \fg
  \oplus \fg^\sh\;,
\end{align}
where the $\fg$-modules on the right hand side appear at heights
$q=-1$, $0$, $0$ and $1$, respectively.
At this point, it is not yet clear that (\ref{S'}) is the full content of $S$ at level $p=0$ since {\apriori}  there might be elements in $\wt S_{(1,1)}$,
$\wt S_{(2,2)}$, {\ldots} that are not contained in $S'=\wt S_{(0,0)}$. We will however see that this is not the case. It suffices to show that
$[f_1,\wt S_{(1,0)}]\subset \wt S_{(0,0)}$.

\subsubsection{The subspace $S_1$}

Before studying the full subspace $S_1$ we first study $\wt S_{(1,0)}\subseteq S_1$ in order to show that $[f_1,\wt S_{(1,0)}]\subseteq \wt S_{(0,0)}$, which implies that
$S_1=\wt S_{(1,0)}$.

The subspace $\wt S_{(1,0)}$ of $\wt S$ is spanned by all elements 
of the form
\begin{align}
[s_1,[s_2,\ldots,[s_{N-1},[s_{N},e_1]]\cdots]] \label{multibracket'}
\end{align}
where $s_1,\ldots,s_{N-1} \in S'$ for some $N\geq0$. 
It follows from the relations in $S'$ that any such expression can be written
as a sum of other ones, which are ``normal-ordered'' in the following sense:
\begin{align}
s_1,\ldots,s_P &\in \fg\;,\nn\\
s_{P+1},\ldots,s_{P+Q} &\in \fg^\sh\;,\nn\\
s_{P+Q+1},\ldots,s_{P+Q+R} &=e_0\;, \label{assumptions}
\end{align}
where $P,Q,R \geq0$ and $P+Q+R=N$. We note that any such nonzero expression is antisymmetric in 
$s_{P+1},\ldots,s_{P+Q}$ since $[\fg^\sh,\fg^\sh]=0$ and we may also without loss of generality assume that it is symmetric in
$s_1,\ldots,s_P$. Furthermore,
because of the relation $[e_0,[e_0,e_1]]=0$ 
we can assume $R$ to be either 
$0$ or $1$.
We will see that this
holds also for $Q$, and we will also restrict the $\lambda$-levels of the elements in $\fg$ and $\fg^\sh$.
First we will show that
$[\fg_{(\leq 1)}{}^\sh,e_1]=0$.

Consider $[x^\sh,e_1]$, where $x \in \fg$.
From the relations $[h_{i}{}^\sh,e_1]=-[f_{0i},e_1]=0$ we know that this is zero if $x$ belongs to the Cartan subalgebra $\mathfrak{h}$ of $\fg$. If $x$ is a root vector $e_\alpha$ of a root $\alpha$ such that $(\alpha,\lambda)\leq0$,
then $[e_\alpha,e_1]=0$ and
\begin{align} \label{e1gnopos}
[x^\sh,e_1]=\frac1{(\alpha,\beta)}[[e_\alpha,f_{0\beta}],e_1]=\frac1{(\alpha,\beta)}\big([e_\alpha,[f_{0\beta},e_1]]-[f_{0\beta},[e_\alpha,e_1]]\big)=0\;,
\end{align}
for some $\beta\in\mathfrak{h}^\ast$ such that $(\alpha,\beta)\neq0$. Thus $[\fg_{(\leq 0)}{}^\sh,e_1]=0$. 
If $x$ is a root vector $e_\alpha$ of a root $\alpha$ such that $(\alpha,\lambda)=1$, then $[e_\alpha,e_1]\neq0$, but still
$[e_\alpha,[e_\alpha,e_1]]=0$.
This implies
\begin{align}
(\ad e_\alpha)^2 (\ad e_1)-2(\ad e_\alpha)(\ad e_1)(\ad e_\alpha)+(\ad e_1)(\ad e_\alpha)^2=0\;,
\end{align}
and then, using (\ref{fundamentalrelation2}),
\begin{align}
[e_1,e_\alpha{}^\sh]&=\frac12[e_1,[e_\alpha,f_{0\alpha}]]\nn\\
&=\frac14[e_1,[e_\alpha,[e_\alpha,[e_{-\alpha},f_{0\alpha}]]]]\nn\\
&=\frac12[e_\alpha,[e_1,[e_\alpha,[e_{-\alpha},f_{0\alpha}]]]]-\frac14[e_\alpha,[e_\alpha,[e_1,[e_{-\alpha},f_{0\alpha}]]]]\nn\\
&=[e_\alpha,[e_1,f_{0\alpha}]]-\frac14[e_\alpha,[e_\alpha,[e_{-\alpha},[e_1,f_{0\alpha}]]]]=0\;. \label{e1esh}
\end{align}
Thus $[\fg_{(\leq 1)}{}^\sh,e_1]=0$.
Since $[\fg_{(\leq0)},e_1]=0$ and $[\fg_{(\ell)},\fg_{(\ell')}{}^\sh]=\fg_{(\ell+\ell')}{}^\sh$ we can now refine
(\ref{assumptions}) to
\begin{align}
s_1,\ldots,s_P &\in \fg_{(\geq1)}\;,\nn\\
s_{P+1},\ldots,s_{P+Q} &\in \fg_{(\geq2)}{}^\sh\;,\nn\\
s_{P+Q+1},\ldots,s_{P+Q+R} &=e_0\;. \label{assumptions2}
\end{align}

Next we will show that $[f_1,\fg_{(\geq1)}{}^\sh]=0$.
Acting on (\ref{e1esh}), where $(\alpha,\lambda)=1$, twice with $f_1$ we get
\begin{align}
0=[f_1,f_1,e_1,e_\alpha{}^\sh]&=-[f_1,h_1,e_\alpha{}^\sh]-[h_1,f_1,e_\alpha{}^\sh]+[e_1,f_1,f_1,e_\alpha{}^\sh]\nn\\
&=(2((\lambda,\alpha)-1)+2)[f_1,e_\alpha{}^\sh]\nn\\
&=2(\lambda,\alpha)[f_1,e_\alpha{}^\sh]\;,
\end{align}
where we have used that
\begin{align}
[f_1,f_1,e_\alpha{}^\sh]=\frac1{(\alpha,\beta)}[f_1,f_1,e_\alpha,f_{0\beta}]=\frac1{(\alpha,\beta)}[e_\alpha,f_1,f_1,f_{0\beta}]=0\;
\end{align}
for some $\beta$ such that $(\alpha,\beta)\neq0$.
Thus $[f_1,\fg_{(1)}{}^\sh]=0$.
If $x\in \fg_{(\ell)}$ for $\ell \geq2$,
then $x^\sh$ is a sum of terms
$[x_{1},\ldots,{x_\ell},f_{0\gamma}]$
where $x_{1},\ldots,{x_\ell} \in \fg_{(1)}$ and
\begin{align}
[f_1,x_{1},\ldots,{x_\ell},f_{0\gamma}]=[x_{1},\ldots,{x_{\ell-1}},f_1,{x_\ell},f_{0\gamma}]=0\;.
\end{align}
Thus $[f_1,\fg_{(\geq1)}{}^\sh]=0$. Since in particular $[f_1,\fg_{(\geq2)}{}^\sh]=0$, and also $[f_1,\fg_{(\geq1)}]=0$, we get
\begin{align} \label{f1e1cancellation}
[f_1,s_1,\,\ldots,s_{N},e_1]=[s_1,\,\ldots,s_{N},f_1,e_1]=-[s_1,\,\ldots,s_{N},h_1] \in S'\;
\end{align}
when we act with $f_1$ on (\ref{multibracket'}), assuming (\ref{assumptions2}). We conclude that $[f_1,\wt S_{(1,0)}]\subseteq \wt S_{(0,0)}$
and it follows that
\begin{align}
\wt S_{-1}&=\wt S_{(0,1)}\;, & \wt S_{0}&=\wt S_{(0,0)}\;, & \wt S_{1}&=\wt S_{(1,0)}\;.
\end{align}

In the same way as in (\ref{f1e1cancellation}), for any $x,y\in \fg_{(\geq 2)}$, we get
\begin{align}
[f_1,s_1,\,\ldots,s_{N},x^\sh,y^\sh,e_1]=-[s_1,\,\ldots,s_{N},x^\sh,y^\sh,h_1]\;.
\end{align} 
This is proportional to $[s_1,\,\ldots,s_{N},x^\sh,y^\sh]$, which is zero, since $[x^\sh,y^\sh]$ is. It follows that
$[x^\sh,[y^\sh,e_1]]$ generates an ideal of $\wt S$ that is contained in $\bigoplus_{p\geq1} {\wt S}_p$, and then it must be zero in $S$ since
$S$ is obtained from $\wt S$ by factoring out the maximal ideal that intersects ${\wt S}_0$ trivially.
Thus $[\fg^\sh,[\fg^\sh,e_1]]=0$ in $S$. 
Furthermore, since
\begin{align}
[x^\sh,y^\sh,e_0,e_1]&=-[x^\sh,y,e_1]-[x^\sh,e_0,y^\sh,e_1]\nn\\
&=-[x^\sh,y,e_1]+[x,y^\sh,e_1]+[e_0,x^\sh,y^\sh,e_1]\nn\\
&=-[[x,y]^\sh,e_1]-[y,x^\sh,e_1]+[x,y^\sh,e_1]=0\;.
\end{align}
we have $[\fg^\sh,\fg^\sh,e_0,e_1]\subseteq \fg([\fg^\sh,e_1])$ and $[\fg^\sh,\fg^\sh,\fg^\sh,e_0,e_1]=0$. We get
\begin{align}
S_1&=\fg([e_0,e_1])+\fg(e_1)+\fg([\fg^\sh,e_0,e_1])+\fg([\fg^\sh,e_1])\nn\\
&=\fg(e_1{}^\fl)+\fg(e_1)+\fg([\fg^\sh,e_1{}^\fl])+\fg([\fg^\sh,e_1])\;.
\end{align}
Here we can replace
$\fg([\fg^\sh,e_1{}^\fl])$ by $\fg([\fg^\sh,e_1]^\fl)$.
We then get
\begin{align} \label{S'e1}
S_1 &=  \fg(e_1{}^\fl)+\fg(e_1)+ 
\fg\big([\fg
^\sh,e_1]^\fl\big) +  
\fg\big([\fg
^\sh,e_1]\big)\;,
\end{align}
where $\fg$ can be replaced by $\fg_{(\geq1)}$ and $\fg^\sh$ can be replaced by $\fg_{(\geq2)}{}^\sh$.
We will see later that this sum of $\fg$-modules is direct. The
$\fg$-modules on the right hand side appear at heights
$q=0$, $1$, $1$ and $2$, respectively.

\subsubsection{The subspace $S_{-1}$}

We now turn to level $p=-1$ and the subspace $S_{-1}=S'(f_1)$. 
It is spanned by all elements 
of the form
\begin{align}
[s_1,\,\ldots,s_{N},f_1] \label{multibracket3}
\end{align}
where $s_1,\ldots,s_{N-1} \in S'$ for some $N\geq0$,
and, according to what we have already shown, we may assume
\begin{align}
s_1,\ldots,s_P &\in \fg_{(\leq-1)}\;,\nn\\
s_{P+1},\ldots,s_{P+Q} &\in \fg_{(\leq 0)}{}^\sh\;, 
\label{assumptions3}
\end{align}
where $P,Q\geq0$ and $P+Q=N$. 
We will show that $[\fg^\sh,\fg^\sh,\fg^\sh,f_1]=0$, which means that $Q\leq2$ in any nonzero expression of the form (\ref{multibracket3}).
According to the results above,
it is sufficient to show that $[e_1,S'([\fg^\sh,\fg^\sh,\fg^\sh,f_1])]=0$ since that implies that $S'([\fg^\sh,\fg^\sh,\fg^\sh,f_1])$ generates an ideal of $S$ contained in $\bigoplus_{p\leq-1} S_p$, which must be trivial. The vector space $S'([\fg^\sh,\fg^\sh,\fg^\sh,f_1])$ is spanned by elements of the form
\begin{align} \label{e1S'xyzf1}
[e_1,s_1,\,\ldots,s_{N},x^\sh,y^\sh,z^\sh,f_1]
\end{align}
where $x,y,z\in \fg$ and $s_1,\,\ldots,s_{N}$ are elements in $S'$ that we can assume satisfy (\ref{assumptions2}) with $R=0,1$. If $R=0$ (that is, if $s_N\neq e_0$),
then we may as well assume that $s_1,\,\ldots,s_{N}$ satisfy (\ref{assumptions3}). Since $[e_1,\fg_{(\leq1)}{}^\sh]=0$
(in particular $[e_1,\fg_{(\leq0)}{}^\sh]=0$, see (\ref{e1gnopos})) and $[e_1,\fg_{(\leq-1)}]=0$ we then get
\begin{align} \label{e1S'xyzf1'}
[e_1,s_1,\,\ldots,s_{N},x^\sh,y^\sh,z^\sh,f_1]&=[s_1,\,\ldots,s_{N},x^\sh,y^\sh,z^\sh,e_1,f_1]\nn\\
&=[s_1,\,\ldots,s_{N},x^\sh,y^\sh,z^\sh,h_1]=0\;.
\end{align}
If $R=1$ (that is, if $s_N= e_0$), then the expression (\ref{e1S'xyzf1}) is equal to
\begin{align}
[e_1,s_1,\,\ldots,s_{N-1},e_0,x^\sh,y^\sh,z^\sh,f_1]&=
-[e_1,s_1,\,\ldots,s_{N-1},x,y^\sh,z^\sh,f_1]\nn\\&\quad\,+[e_1,s_1,\,\ldots,s_{N-1},x^\sh,y,z^\sh,f_1]\nn\\&\quad\,
-[e_1,s_1,\,\ldots,s_{N-1},x^\sh,y^\sh,z,f_1]\;,
\end{align}
which in turn can be written as a sum of terms of the form
\begin{align}
[e_1,s_1,\,\ldots,s_{N},x^\sh,y^\sh,f_1]
\end{align}
where $x,y \in \fg$ and $s_1,\,\ldots,s_{N}$ satisfy (\ref{assumptions3}). This can be shown to be zero in the same way as 
$[e_1,s_1,\,\ldots,s_{N},x^\sh,y^\sh,z^\sh,f_1]$ in (\ref{e1S'xyzf1'}).

Thus we have $[\fg^\sh,\fg^\sh,\fg^\sh,f_1]=0$, and it follows that
\begin{align}
S_{-1}=\fg(f_1)+\fg([\fg^\sh,f_1])+\fg([\fg^\sh,\fg^\sh,f_1])\;.
\end{align}
As we will see, it is convenient to rewrite this sum of $\mathfrak{g}$-modules.
First, since
\begin{align}
[x^\sh,y^\sh,f_1]&=-[x^\sh,y^\sh,e_0,f_1{}^\sh]\nn\\
&=[x^\sh,y,f_1{}^\sh]-[x,y^\sh,f_1{}^\sh]
-[e_0,x^\sh,y^\sh,f_1{}^\sh]\nn\\
&=[[x,y]^\sh,f_1{}^\sh]+[y,x^\sh,f_1{}^\sh]-[x,y^\sh,f_1{}^\sh]\;,
\end{align}
we have
$\fg([\fg^\sh,\fg^\sh,f_1])=\fg([\fg^\sh,f_1{}^\sh])$ .
Second, it will turn out to be convenient to write $\fg([\fg^\sh,f_1])$ as a sum of the two submodules $\fg(f_1{}^\sh)$ and
$\fg([\fg^\sh,f_1{}^\sh]^\fl)$. We thus arrive at
\begin{align}
S_{-1}&= 
\fg(f_1)+\fg(f_1{}^\sh) 
+\fg \big([\fg^\sh,f_1{}^\sh]^\fl\big)
+ 
\fg \big([\fg
^\sh,f_1{}^\sh]\big)
\;, \label{S'f1}
\end{align}
where $\fg$ can be replaced by $\fg_{(\leq-1)}$ and $\fg^\sh$ can be replaced by $\fg_{(\leq0)}{}^\sh$.
The $\fg$-modules on the right hand side appear at heights
$q=-1$, $0$, $0$ and $1$, respectively.

In the above derivation of the content of $S_{\pm1}$ in terms of
$\fg$-modules, we have relied on the definition of $S$ as the
superalgebra obtained by factoring out the maximal ideal in $\wt S$
intersecting $\wt S_0$ trivially. 
We know that in many cases \cite{Carbone:2018xqq}, the ideal contains a part generated by $[f_{0i},[f_{0j},f_1]]$
for all $i,j$ such that $\lambda_i=\lambda_j=0$. We also know that in some cases there is an additional part generated by elements at positive levels $p\geq 2$,
see Section \ref{IdealKSubSec}.
Although we have not been able to derive the content of $S_{\pm1}$ in terms of
$\fg$-modules using only the defining relations (and $[f_{0i},[f_{0j},f_1]]$ for $\lambda_i=\lambda_j=0$)
we have no proof that it is impossible. This possibility of course does not affect the
results \eqref{S'e1} and \eqref{S'f1}.

\section{The tensor product $R(\lambda) \otimes {\bf adj}$
  \label{TensorProductSection}}

We will now determine the $\fg$ modules that appear in the local part of $S$, that is, on the right hand sides of (\ref{S'}), (\ref{S'e1}) and
(\ref{S'f1}).
At $p=0$ we already know that $\fg^\sh$ is an adjoint $\fg$ module, and that $e_0$ and $\wt k$ span two singlets.
At $p=1$ it is easy to see that
$\fg(e_1{}^\fl)$ and $\fg(e_1)$ are lowest-weight modules with lowest weights $-\lambda$,
\begin{align}
\fg(e_1{}^\fl) \simeq \fg(e_1) \simeq R(-\lambda)\;.
\end{align}
Likewise, at $p=-1$ it is easy to see that
$\fg(f_1)$ and $\fg(f_1{}^\sh)$ are highest-weight modules with highest weights $\lambda$,
\begin{align}
\fg(f_1) \simeq \fg(f_1{}^\sh) \simeq R(\lambda)\;.
\end{align}
It remains to determine the modules
\begin{align}
\fg \big([\fg{}^\sh,e_1]^\fl\big)\simeq
\fg \big([\fg{}^\sh,e_1]\big) \label{e-modul}
\end{align}
at $p=1$ and 
\begin{align}
\fg \big([\fg{}^\sh,f_1{}^\sh]^\fl\big)\simeq
\fg \big([\fg{}^\sh,f_1{}^\sh]\big)  \label{f-modul}
\end{align}
at $p=-1$.
These modules must be contained in the tensor products $R(-\lambda) \otimes {\bf adj}$ and $R(\lambda) \otimes {\bf adj}$, respectively.
We will therefore in this section study the tensor product $R(\lambda) \otimes {\bf adj}$ and its decomposition into a direct sum of irreducible
submodules. (It is of course sufficient to study one of the two tensor products in detail.)

We thus consider the tensor product $R(\lambda)\otimes\adj$, where $\lambda$
is an arbitrary dominant integral weight. Clearly, all
irreducible representations occurring in the tensor products with
non-zero multiplicity are $R(\lambda+\gamma)$, where $\gamma$ lies in
the root lattice.

Denote the multiplicity of $R(\nu)$ in $R(\mu)\otimes R(\lambda)$ by
$\mult(R(\mu)\otimes R(\lambda),R(\nu))$. The multiplicity formula of
Parthasarathy, Ranga Rao and Varadarajan (PRV) \cite{ParthasarathyEtAl}
reads
\begin{align}
  \mult(R(\lambda)\otimes R(\mu),R(\nu))
  =\dim\{v\in R(\mu)_{\nu-\lambda}:e_i^{\lambda_i+1}v=0 \ \, \textrm{for all}\ \,  i\}\;,
  \label{PRBFormula}
\end{align}
where $R(\mu)_\nu$ denotes the subspace of $R(\mu)$ at weight
$\nu$. The r\^oles of $\lambda$ and  $\mu$ can of course be
interchanged in the formula. A state $v\in R(\mu)_{\nu-\lambda}$ such
that $e_i^{\lambda_i+1}v=0$ for all $i$ will be called PRV state
below. A PRV state is in general not a highest weight state for the
corresponding irreducible representation in the tensor product, but
always a part of it.
Applied to the tensor product under consideration, we  get
\begin{align}
  \mult(R(\lambda)\otimes\adj ,R(\lambda+\gamma))
  =\dim\{v\in\adj_\gamma:(\ad e_i)^{\lambda_i+1}v=0 \ \, \textrm{for
    all}\ \,  i\}\;. 
\end{align}
This shows that the multiplicity can only be non-zero when
$\gamma\in\Gamma\cupp\{0\}$, where $\Gamma$ is the root space of $\fg$.
It also immediately follows that non-zero multiplicities of
$R(\lambda+\gamma)$, $\gamma\neq0$, equal $1$, due to the
non-degeneracy of the root decomposition of the (finite-dimensional)
Lie algebra $\fg$.

First consider PRV states $v$ for $\gamma\neq0$.
For $i$ such that $\lambda_i=0$, we need $e_iv=0$, which means that
$\gamma$ is a highest root at some $\lambda$-level. For $i$ such that
$\lambda_i\neq0$, we have $[e_i,[e_i,v]]=0$ for such roots, with the
only exception $\ell=-1$, $v=f_i$ for an $i$ with $\lambda_i=1$.
When $\gamma=0$, we need elements in the Cartan algebra, which are
annihilated by all $e_i$ for which $\lambda_i=0$. These are linear
combinations of $h_j$ for $\lambda_j\neq0$, namely the fundamental weights $\Lambda_j$, and they trivially satisfy the
remaining conditions.

Thus, we have shown that
\begin{align}\label{RlambdaAdjRepr}
  R(\lambda)\otimes\adj=NR(\lambda)\oplus
  \bigoplus\limits_{\substack{\ell=-(\lambda,\theta)\\ \ell\neq-1}}^{(\lambda,\theta)}
  \bigoplus\limits_{\gamma^{(\ell)}\in H_\ell}
  R(\lambda+\gamma^{(\ell)})\;,
\end{align}
where $H_\ell$ is the set of highest roots at $\lambda$-level
$\ell$, and $N$ is the number of non-zero $\lambda_i$ ($N=1$ for
$\lambda$ a multiple of a fundamental weight).

At a given $\ell$, there
may be several roots in $H_\ell$.
All $\lambda$-levels from $-(\lambda,\theta)$ to
$(\lambda,\theta)$ occur, except $\ell=-1$, assuming $\lambda$
is not a multiple of a smaller integral dominant weight. If
$\lambda=n\lambda'$, the relevant $\lambda$-levels are $-n(\lambda',\theta),-n((\lambda',\theta)+1),\ldots,n(\lambda',\theta)$.

We introduce the notation
\begin{align}
R_{(\ell)}=\bigoplus\limits_{\gamma^{(\ell)}\in H_\ell}
  R(\lambda+\gamma^{(\ell)})
\end{align}
for $\ell\neq0,-1$ and
\begin{align}
R_{(0)}&=NR(\lambda)\oplus\bigoplus\limits_{\gamma^{(0)}\in H_0}
  R(\lambda+\gamma^{(0)}) \;, \qquad R_{(-1)}=\nullrep
\end{align}
so that
$R(\lambda)\otimes\adj=\bigoplus_{\ell=-(\lambda,\theta)}^{(\lambda,\theta)}R_{(\ell)}$.
We will show that 
\begin{align} \label{modullikhet}
\fg (v_\lambda \otimes \fg_{(\geq 1)}) = \bigoplus_{\ell=1}^{(\lambda,\theta)} R_{(\ell)}\;,
\end{align}
where $v_\lambda$ is a lowest weight state in $R(\lambda)$.
Any element in $v_\lambda \otimes \fg_{(\geq 1)}$ must belong to the module on the right hand side of (\ref{modullikhet}), since the complementary submodule
$\bigoplus_{\ell=-(\lambda,\theta)}^{0} R_{(\ell)}$ of $R(\lambda) \otimes \fg$ is spanned by states of lower weights. Thus the left hand side of (\ref{modullikhet})
is contained in the module on the right hand side. Conversely, the highest weight state in any submodule $R_{(\ell)}$ 
must be a linear combination of $v_\lambda \otimes e_{\gamma^{(\ell)}}$ and elements in
\begin{align}
\fg_- (v_\lambda) \otimes \fg_+\big(e_{\gamma^{(\ell)}}\big) \subseteq \fg_- \big(v_\lambda \otimes \fg_+(e_{\gamma^{(\ell)}})\big)+
v_\lambda \otimes \fg_-\big(\fg_+(e_{\gamma^{(\ell)}})\big)
\end{align}
of weight $\lambda+\gamma^{(\ell)}$, where $\fg_\pm$ denote the Borel subalgebras of $\fg$ spanned by $\{e_i\}$ and $\{f_i\}$, respectively. But the only elements in $v_\lambda \otimes \fg_-\big(\fg_+(e_{\gamma^{(\ell)}}\big)$ of weight $\lambda+\gamma^{(\ell)}$
are multiples of $v_\lambda \otimes e_{\gamma^{(\ell)}}$. Thus the highest weight state in any module $R_{(\ell)}$ belongs to
$\fg (v_\lambda \otimes \fg_{(\geq \ell)})$ and it follows that the module on the right hand side of (\ref{modullikhet}) is contained in the module on the left hand side. We conclude that (\ref{modullikhet}) holds.

Let us now return to the modules in (\ref{e-modul}) and (\ref{f-modul}). 
We have seen that $[x^\sh,f_1{}^\sh]$ is zero for $x\in\fg_{(\geq1)}$. 
Also, if $x \in \mathfrak{h}_\lambda$, where $\mathfrak{h}_\lambda$ is the subspace of $\mathfrak{h}$ spanned by $h_\lambda$,
then we have $[x^\sh,f_1{}^\sh]=0$ since $2[f_{0\lambda},[f_{0\lambda},f_1]]=[[f_{0\lambda},f_{0\lambda}],f_1]=0$.
On the other hand, if $x$ is a root vector $x\in\fg_{(0)}$ or $x\in\fg_{(\leq-2)}$,
then $[x^\sh,f_1{}^\sh]$ is nonzero. This can be seen by acting with first $e_0$ and then $e_1$. We then get
\begin{align} \label{egenvarde}
[e_1,e_0,x^\sh,f_1{}^\sh]=\Bigg(\bigg(1-\frac1{(\lambda,\lambda)}\bigg)\ell(x)-1\Bigg)x^\sh\,,
\end{align}
where $\ell(x)$ is the $\lambda$-level of $x$. 
If $\ell(x)\leq-2$ for some nonzero $x \in\fg$
then $(\lambda,\lambda)\geq1$
since the only case where $(\lambda,\lambda)<1$ is $\fg=A_r$, $\lambda=\Lambda_1$ (or $\lambda=\Lambda_r$), which leads to a 3-grading 
$\fg=\fg_{(-1)}\oplus\fg_{(0)}\oplus\fg_{(1)}$ (this can be checked by inspecting the inverse Cartan matrices for simply laced Lie algebras $\fg$)
and then
\begin{align}
\bigg(1-\frac1{(\lambda,\lambda)}\bigg)\ell(x)-1 \leq -1\;.
\end{align}
Also for $x \in \mathfrak{h}'$, where $\mathfrak{h}'$ is a subspace of $\mathfrak{h}$ such that $\mathfrak{h}=\mathfrak{h}_\lambda \oplus\mathfrak{h}'$
(if $N>1$) it is easy to check that $[x^\sh,f_1{}^\sh]\neq0$.
It follows that 
\begin{align}
\fg \big([\fg{}^\sh,f_1{}^\sh]^\fl\big)\simeq\fg \big([{\fg_{(\leq0)}}^\sh,f_1{}^\sh]\big) &\simeq 
\frac{R{(\lambda)} \otimes \fg}{\fg (v_\lambda \otimes \fg_{(\geq 1)}) \oplus \fg (v_\lambda \otimes \mathfrak{h}_\lambda)}\nn\\
&\simeq\bigoplus_{\ell=-(\lambda,\theta)}^{0} R_{(\ell)}\ominus
R(\lambda)\;.
\label{PhiModules}
\end{align}
Similarly, at level $p=1$ we find that
\begin{align}
\fg \big([\fg{}^\sh,e_1]^\fl\big)\simeq
\fg \big([\fg{}^\sh,e_1]\big) \simeq 
\frac{R{(-\lambda)} \otimes \fg}{\fg (u_{-\lambda} \otimes \fg_{(\leq
    1)})}
\simeq\bigoplus_{\ell=-(\lambda,\theta)}^{-2} \overline{R_{(\ell)}}\;,
\end{align}
where $u_{-\lambda}$ is a lowest weight state in $R(-\lambda)$. This
is the representation $\wt R_1$. 

\section{Construction from $\fg$-representations\label{gReprSection}}

\subsection{Local superalgebra in terms of $\fg$-modules}

We have shown that if $S$ is non-trivial, then its local part decomposes into a sum of $\fg$-modules according to (\ref{S'}), (\ref{S'e1}) and
(\ref{S'f1}). In order to show that $S$ indeed
is non-trivial we will now construct a  non-trivial Lie superalgebra
that satisfies the relations (\ref{eigen1})--(\ref{IdealJ2'}) if the generators are identified with certain elements in it.
In this construction we use the fact that there is a $\mathbb{Z}$-graded Lie superalgebra $\scr{U}=\bigoplus_{p\in\mathbb{Z}} \scr U_{p}$
associated to any $\mathbb{Z}_2$-graded vector space $\scr{U}_1$, generalising the universal $\mathbb{Z}$-graded Lie algebra associated to a vector space
\cite{Kantor-graded,Palmkvist:2009qq}. The subspaces $\scr U_{-p}$ for $p\geq 0$ are defined recursively as consisting of all linear maps
$\scr U_1 \to \scr U_{-p+1}$, and the brackets are such that $[A,a]=A(a)$ for $A \in \scr U_{-p}$ ($p\geq 0$)
and $a \in \scr U_1$. In particular, $\scr U_0 = \mathfrak{gl}(\scr U_1)$.
The subalgebra $\bigoplus_{p\geq0} \scr U_{p}$ is freely generated by $\scr U_1$.

In this case, we let $\scr U_1$ be the direct sum of four $\fg$-modules, pairwise isomorphic with an isomorphism $\sh$.
Two of the four $\fg$-modules transform in the representation $R_1=R(-\lambda)$ and are 
denoted by $U$ and $U^\sh$,
respectively. The other two transform in $\wt R_1$ and are 
denoted by $\wt U$ and $\wt U^\sh$,
respectively. 
Thus
\begin{align}
\scr U_1 =  U \oplus U^\sh \oplus  \wt U \oplus\wt U^\sh\;.
\end{align}
According to the discussion in the preceding section, we consider the module $\wt U$
as the quotient
\begin{align}
\wt U = \frac{U\otimes \fg}{\fg(e_1\otimes \fg_{(\leq1)})}\;,
\end{align}
where $e_1$ is a lowest weight state of $U$.
We let $L$ be the natural map 
$U\otimes \fg \to \wt U$, so that $L(u \otimes x)=0$ if and only
if $x \in \fg_{(\leq1)}$.

Since $\scr U_1$ is a $\fg$-module, we can consider $\fg$ as a subalgebra of $\scr U_0=\mathfrak{gl}({\scr U}_1)$. We then define
an odd subspace $\fg^\sh$ of $\scr U_0$ isomorphic to $\fg$, an odd
element $e_0 \in \scr U_0$ and an even element $\wt k \in \scr U_0$ by
\begin{align}
[x^\sh, L(u \otimes y)^\sh]&=0\;, \nn\\
[x^\sh, L(u \otimes y)]&=  [x,L(u \otimes y)]^\sh=L([x,u]\otimes y)^\sh+L(u \otimes [x,y])^\sh\;, \nn\\
[x^\sh, u^\sh]&= -L(u\otimes x)^\sh\;,\nn\\
[x^\sh, u]&=-[x,u]^\sh-L(u\otimes x) \;, \label{x^sh-verkan}
\end{align}
\begin{align}
[e_0,L(u\otimes x)^\sh]&=-L(u\otimes x)\;,\nn\\
[e_0,L(u\otimes x)]&=0\;,\nn\\
[e_0,u^\sh]&=u\;,\nn\\
[e_0,u]&=0\;,
\end{align}
\begin{align}
[\wt k,L(u\otimes x)^\sh]&=\big(3-(\lambda,\lambda)\big)L(u\otimes x)^\sh\;,\nn\\
[\wt k,L(u\otimes x)]&=\big(2-(\lambda,\lambda)\big)L(u\otimes x)\;,\nn\\
[\wt k,u^\sh]&=\big(2-(\lambda,\lambda)\big)u^\sh\;,\nn\\
[\wt k,u]&=\big(1-(\lambda,\lambda)\big)u\;.
\end{align}
It is then easy to check that the subspace $\langle e_0 \rangle \oplus\langle \wt k \rangle \oplus \fg \oplus \fg^\sh$ of $\scr U_0$ closes under the super-commutator
and thus form a subalgebra.
The brackets are given by
\begin{align}
[x,y^\sh]&=[x,y]^\sh\;, & [e_0,x^\sh]&=-x\;, & [\wt k,x^\sh]&=x^\sh\;,&  [\wt k,e_0]&=-e_0
\end{align}
and $[e_0,e_0]=[\fg^\sh,\fg^\sh]=[e_0,\fg]=[\wt k,\fg]=0$.

We define $e_1^\fl$ by $e_1=(e_1{}^\fl)^\sh$ and define an element $f_1 \in \scr U_{-1}$
recursively by
\begin{align}
[f_1,e_1{}^\fl]&=-e_0\;\nn\\
[f_1,e_1]&=h_\lambda-\wt k\;\nn\\
[f_1,L(e_1\otimes x)]&=
\begin{cases}
(\ell(x)-1)x & \text{ if }\quad \ell(x)\geq 2\;,\\
0 & \text{ if }\quad \ell(x)\leq 1\;,
\end{cases}
\,\nn\\
[f_1,L(e_1\otimes x)^\sh]&=
\begin{cases}
(\ell(x)-1)x^\sh & \text{ if }\quad \ell(x)\geq 2\;,\\
0 & \text{ if }\quad \ell(x)\leq 1\;,
\end{cases}
\end{align}
where $\ell(x)$ is the $\lambda$-level of $x$, and
\begin{align}
[f_1,e_i(u)]&=0\;,\nn\\
[f_1,e_i(u^\sh)]&=[e_i,[f_1,u^\sh]]\;,\nn\\
[f_1,L([e_i,u]\otimes x)]&=[e_i,[f_1,L(u \otimes x)]]-[f_1,L(u \otimes [e_i,x])]\;,\nn\\
[f_1,L([e_i,u]\otimes x)^\sh]&=[e_i,[f_1,L(u \otimes x)^\sh]]-[f_1,L(u \otimes [e_i,x])^\sh]\;.
\end{align}
It is straightforward to show that $f_1$ is well defined and then that all the relations (\ref{eigen1})--(\ref{IdealJ2'}) are satisfied with
$f_{0i}=-h_i{}^\sh$ and $h_1=h_\lambda-\wt k$. Thus there is a surjective isomorphism
from $\wt S$ to the subalgebra of $\scr U$ generated by $f_1 \in \scr U_{-1}$ and ${\scr U}_1$.
It follows that the Lie superalgebra $S$ indeed is non-trivial, the
sums in (\ref{S'}), (\ref{S'e1}) and (\ref{S'f1}) are direct, and the $\fg$-modules that appear can be decomposed into highest and lowest weight modules according to the discussion in
the preceding section. 

\subsection{Covariant description}

Let $E_M$ be a basis of $U$. We 
set $L_{\alpha M}=-L(E_M \otimes T_\alpha)$, so that $L_{\alpha M}$
is a basis of $\wt U$
(as before,
$T_\alpha$ is a basis of $\fg$). 
Similarly to $S'(e_1) = \scr U_{1}$ at $p=1$,
we decompose the subspace $S'(f_1)$ of $\scr U_{-1}$ at $p=-1$ into $\fg$-modules as
\begin{align}
S'(f_1) =  V \oplus V^\sh \oplus  \wt V \oplus\wt V^\sh\;,
\end{align}
where $V$
transforms in $R(\lambda)$ with lowest weight state $f_1$ and basis $F^M$.
Then we can identify $\wt V$ with the quotient
\begin{align}
\frac{V \otimes \fg}{\fg (f_1 \otimes \fg_{(\geq 1)}) \oplus \fg (f_1 \otimes \mathfrak{h}_\lambda)}\,
\end{align}
and let $\Phi$ be the natural map $V\otimes \fg \to \wt V$.
If we now set $\Phi(F^M \otimes T_\alpha)=-\Phi_\alpha{}^M$, then the brackets in $S$ involving $\fg^\sh$ and the modules at level $p=\pm1$ 
can be
written on tensorial form as
\begin{align}
[T_\alpha^\sh, E_{M}]&=t_{\alpha M}{}^N E_N^\sh+L_{\alpha M}\nn\;,\\
[T_\alpha^\sh, E_{M}^\sh]&=L_{\alpha M}^\sh\;,\nn\\
[T_\alpha^\sh, L_{\beta M}]&=[T_\alpha, L_{\beta M}]^\sh=f_{\alpha\beta}{}^\gamma L_{\gamma M}^\sh-t_{\alpha M}{}^N L_{\beta N}^\sh\;,\nn\\
[T_\alpha^\sh, L_{\beta M}^\sh]&=0\;,
\label{TshELBrackets}
\end{align}
\begin{align}
[T_\alpha^\sh,H^{\fl M}]&=t_{\alpha N}{}^MH^N+\Phi_\alpha{}^M\nn\;,\\
[T_\alpha^\sh,H^M]&=-\Phi^\sh_\alpha{}^M\nn\;,\\
[T_\alpha^\sh,\Phi_\beta{}^M]&=-[T_\alpha,\Phi_\beta{}^M]^\sh
=-f_{\alpha\beta}{}^\gamma\Phi^\sh_\gamma{}^M-t_{\alpha N}{}^M\Phi^\sh_\alpha{}^N\nn\;,\\
[T_\alpha^\sh,\Phi^\sh_\beta{}^M]&=0\;,
\label{TshHPhiBrackets}
\end{align}
\begin{align}
[H^{\fl M},E_N]&=- \delta_N{}^M e_0\;,\nn\\
[H^{\fl M},E_N^\sh]&=t_{\alpha N}{}^M T^\alpha - \delta_N{}^M \wt k\;,\nn\\
[H^M,E_N]&=-\left(1-\frac1{(\lambda,\lambda)}\right)t^\alpha{}_N{}^MT_\alpha
  +\delta_N^M\tk\;,\nn\\
[H^M,E^\sh_N]&=\frac1{(\lambda,\lambda)}t^\alpha{}_N{}^MT^\sh_\alpha\;,
\label{HEBrackets}
\end{align}
\begin{align}
    [H^{\fl M},L_{\alpha N}]&=-\ell_{\alpha N}{}^{\beta M}T_\beta\;,\nn\\
[H^{\fl M},L_{\alpha N}^\sh]&=-\ell_{\alpha N}{}^{\beta M}T_\beta^\sh\;,\nn\\
[H^M,L_{\alpha N}]&=-\ell_{\alpha N}{}^{\beta M}T^\sh_\beta\;,\nn\\
[H^M,L^\sh_{\alpha N}]&=0\;,
\label{HLBrackets}
\end{align}
\begin{align}
[\Phi_\alpha{}^M,E_N]&=\varphi^\beta{}_{N,\alpha}{}^MT_\beta\;,\nn\\
[\Phi_\alpha{}^M,E_N^\sh]&=\varphi^\beta{}_{N,\alpha}{}^MT^\sh_\beta\;,\nn\\
[\Phi^\sh_\alpha{}^M,E_N]&=-\varphi^\beta{}_{N,\alpha}{}^MT^\sh_\beta\;,\nn\\
[\Phi^\sh_\alpha{}^M,E^\sh_N]&=0\;,
\label{PhiEBrackets}
\end{align}
\begin{align}
  [L_{\alpha M},\Phi_\beta{}^N]=[L_{\alpha M},\Phi^\sh_\beta{}^N]=
  [L^\sh_{\alpha M},\Phi_\beta{}^N]=[L^\sh_{\alpha M},\Phi^\sh_\beta{}^N]=0\;,
\end{align}
for some invariant tensors $\ell_{\alpha N}{}^{\beta M}$ and
$\varphi^\alpha{}_{M,\beta}{}^N$. These tensors will be some linear
combinations of projectors on the modules appearing in $L$ and
$\Phi$. The coefficients in these linear combinations are completely
determined. One may think of $L$ and $\Phi$ as defined by their
appearances in the first equations in \eqref{TshELBrackets} and
\eqref{TshHPhiBrackets}. The normalisation is then fixed, and the
tensors $\ell$ and $\varphi$ are determined. As we will see in Section
\ref{ARemarkableIdentity}, they are even seemingly over-determined, and
exist thanks to a peculiar identity.

\begin{table}
  \begin{align*}
  \xymatrix@=.4cm{
    \ar@{-}[]+<2.2em,1em>;[dddd]+<2.2em,-1em>
    \ar@{-}[]+<-0.8cm,-1em>;[rrr]+<1.4cm,-1em>
    &\ar@{-}[]+<4.4em,1em>;[dddd]+<4.4em,-1em> p=-1
    & \ar@{-}[]+<3em,1em>;[dddd]+<3em,-1em> p=0 &p=1\\
q=2&&&L^\sh_{\alpha M}\\
q=1&\Phi_{\alpha}^{\sh M}\quad G^{\sh M}&f_{00}\quad T_\alpha^\sh& E_M^\sh\quad L_{\alpha M}
       \\ 
q=0&F^M\quad\Phi_{\alpha}{}^M\quad G^M
&k\quad T_\alpha\quad\tk
           & E_M\\ 
q=-1 & F^{\fl M}& {e_0} 
  }
\end{align*}
  \caption{\it Basis elements for $W(\fgrplus)$ at $p=-1,0,1$.}
\label{GeneralWTableBasis}
\end{table}
\begin{table}
  \begin{align*}
  \xymatrix@=.4cm{
    \ar@{-}[]+<2em,1em>;[dddd]+<2em,-1em>
    \ar@{-}[]+<-0.8cm,-1em>;[rrr]+<1.4cm,-1em>
    &\ar@{-}[]+<3.5em,1em>;[dddd]+<3.5em,-1em> p=-1
    & \ar@{-}[]+<2.75em,1em>;[dddd]+<2.75em,-1em> p=0 &p=1\\
q=2&&&L_{\alpha M}^\sh\\
q=1&\Phi^\sh_{\alpha}{}^{M}&T_\alpha^\sh& E_M^\sh\quad\quad L_{\alpha M}
       \\ 
q=0&\Phi_{\alpha}{}^M\quad\quad H^M
&\tk\quad\quad{T_\alpha}
           & E_M\\ 
q=-1 & H^{\fl M}& {e_0} 
  }
\end{align*}
  \caption{\it Basis elements for $S(\fgrplus)$ at $p=-1,0,1$.}
\label{GeneralSTableBasis}
\end{table}
An alternative way of deriving the content of $S_{-1}$ is to note
that the basis elements $E_M$ for $R_1=R(-\lambda)$
have a covariant Serre relation in
$\fg([e_0,e_0])=R(-2\lambda)$, so that the bracket $[E_M,E_N]$ lies in
$R_2=\vee^2R_1\ominus R(-2\lambda)$.
Any element at $(p,q)=(-1,0)$ must respect the ideal in
$R(-2\lambda)$. This allows for the introduction of generators $\Phi_\alpha{}^M$ 
with brackets
$[E_N,\Phi_\alpha{}^M]=\varphi^\beta{}_{N,\alpha}{}^MT_\beta$,
where $\varphi$ is a linear combination of projection operators on the
irreducible modules in $\Phi$. 
They respect the ideal in $R(-2\lambda)$ if
\begin{align}
  t_{\beta\langle M}{}^P\varphi^\beta{}_{N\rangle,\alpha}{}^Q=
  (t_\beta\otimes\varphi^\beta{}_\alpha)_{\langle MN\rangle}{}^{PQ}=0\;,
  \label{EmbeddingTensorCondition}
\end{align}
or equivalently,
$(\varphi^\alpha{}_\beta\otimes t^\beta)_{MN}{}^{\langle PQ\rangle}=0$,
where $\langle MN\rangle$ denotes projection on $R(\pm2\lambda)$.
Eq. \eqref{EmbeddingTensorCondition} is the condition for the
representation of the embedding tensor, or the ``big torsion
representation'' in extended geometry.\footnote{Although we have not
  performed a complete 
  analysis, we have noted that in cases when $\lambda$ is attached to
  a short root, there is typically no solution to this algebraic
  condition.}

Let us check which of the representations in $R(\lambda)\otimes\adj$
that respect the Serre relations.
Consider an irreducible submodule $R(\lambda+\gamma^{(\ell)})$, where
$\gamma^{(\ell)}$ is a highest root at $\lambda$-level $\ell$.
The Serre relations will automatically have vanishing bracket with an
element in this module if $R(\lambda+\gamma^{(\ell)})\otimes
R(-2\lambda)\not\supset R(-\lambda)$, \ie, if
\begin{align}
R(\lambda+\gamma^{(\ell)})\otimes
R(\lambda)\not\supset R(2\lambda)\;.
\end{align}
Applying the PRV formula \eqref{PRBFormula} for the multiplicity of
$R(2\lambda)$ in the tensor product on the left hand side, we obtain
\begin{align}
&\mult(R(\lambda+\gamma^{(\ell)})\otimes R(\lambda),R(2\lambda))\nn\\
  &\quad=\dim\{v\in R(\lambda)_{\lambda-\gamma^{(\ell)}}:e_i^{(\lambda+\gamma^{(\ell)})_i+1}v=0 \ \, \textrm{for all}\ \,  i\}\;.
\end{align}
This multiplicity is obviously $0$ for $\ell\geq0$, since $R(\lambda)$
does not contain any states with the same or higher $\lambda$-level
than the highest weight state. For the module $R(\lambda)$ we have
$\mult(R(\lambda)\otimes R(\lambda),R(2\lambda))=1$.
We arrive at the statement that $\Phi$
respects the Serre relations at level $2$ if it contains the
irreducible modules
\begin{align}
  (N-1)R(\lambda)\oplus\bigoplus\limits_{\gamma\in H_0}R(\lambda+\gamma)
  \oplus\bigoplus\limits_{\ell=2}^{(\lambda,\theta)}
        \bigoplus\limits_{\beta\in L_\ell}R(\lambda-\beta)\;,
\end{align}
where $H_0$ is the set of highest roots at $\lambda$-level $0$ and $L_\ell$ the
set of lowest roots at $\lambda$-level $\ell$.
This is the same sum of irreducible modules as was already shown to 
constitute $\fg \big([\fg{}^\sh,f_1{}^\sh]^\fl\big)$
in eq. \eqref{PhiModules}.

Using the covariant
brackets, one can also check explicitly that $\Phi$ respects the
Serre relations in $\bigoplus_{i:\lambda_i=1}R(-(2\lambda-\alpha_i))$ in
$[E^\sh_M,E^\sh_N]$. The condition becomes
\begin{align}
L^\sh_{\beta\{N}\varphi^\beta{}^{\mathstrut}_{P\},\alpha}{}^M=0\;,
\end{align}
where $\{NP\}$ denotes projection on $\bigoplus_{i:\lambda_i=1}R(\pm(2\lambda-\alpha_i))$.
This is automatically satisfied, since the highest modules in
$\Phi$ and $L$ are $R(\lambda+\gamma_0)$ and $R(\lambda-\beta_2)$,
where $\gamma_0$ is a highest root at level $0$. The tensor product
can not contain $R(2\lambda-\alpha_i)$, where $(\lambda,\alpha_i)=1$,
since $2\lambda-\alpha_i\succ2\lambda+\gamma_0-\beta_2$.

\subsection{A remarkable identity\label{ARemarkableIdentity}}

Consider the Jacobi identity between $T^\sh_\alpha$, $E_M$
and $H^{\fl N}$. This turns out to be the only non-trivial Jacobi
identity within  the local superalgebra at $p=-1,0,1$, in the sense
that all others can be obtained from it by raising and lowering
operations.
A short calculation leads to the necessary and sufficient condition
for this Jacobi identity to be fulfilled:
\begin{align}  
  \varphi^\beta{}_{M,\alpha}{}^N-\ell_{\alpha M}{}^{\beta N}
  &=\delta_\alpha^\beta\delta_M^N
  -f_\alpha{}^{\beta\gamma}t_{\gamma M}{}^N
  -\frac1{(\lambda,\lambda)}(t^\beta t_\alpha)_M{}^N
  \equiv Q_{\alpha M}{}^{\beta N}\;,
  \label{PhiLRelation}
\end{align}
\ie,
\begin{align}
  \varphi^ \beta{}_\alpha-\ell_\alpha{}^\beta=\delta_\alpha^\beta
  -f_\alpha{}^{\beta\gamma}t_\gamma-\frac1{(\lambda,\lambda)}t^\beta
  t_\alpha
  \equiv Q_\alpha{}^\beta\;.
  \label{PhiLRelation2}
\end{align}
If we now make use of the algebraic condition
\eqref{EmbeddingTensorCondition} on $\varphi$, the part of
this relation only  involving $\ell$ becomes
\begin{align}
  \ell_{\beta M}{}^{\alpha\langle P}t^\beta{}_N{}^{Q\rangle}
  &=f^\alpha{}_{\beta\gamma}t^\beta{}_M{}^{\langle P}t^\gamma{}_N{}^{Q\rangle}
  +t^\alpha{}_M{}^{\langle P}\delta_N^{Q\rangle}
  -\delta_M^{\langle P}t^\alpha{}_N{}^{Q\rangle}\nn\\
  &=(f^\alpha{}_{\beta\gamma}t^\beta\otimes t^\gamma+t^\alpha\otimes\id
  -\id\otimes t^\alpha)_{MN}{}^{\langle PQ\rangle}
  \;.
  \label{EllTIsS}
\end{align}
The right hand side is recognised as the ``$S$ tensor'' of ref.
\cite{Cederwall:2017fjm}. (There, a non-vanishing $S$ tensor was shown
to be equivalent to the presence of ancillary transformations in the
commutator of two generalised diffeomorphisms.
Here, it is related to the existence of a module $\tR_1$.
See also ref. \cite{CederwallPalmkvistTHAII}.)
We thus have
\begin{align}
  (\ell_\beta{}^\alpha\otimes t^\beta)_{MN}{}^{\langle PQ\rangle}
  =S^\alpha{}_{MN}{}^{PQ}\;.
  \label{EllTIsS2}
\end{align}
The tensor $S$ is antisymmetric in its lower
indices. In addition, it satisfies
$S_{\{MN\}}{}^{PQ}=0$, thanks to the identity
\begin{align}
  S^\alpha{}_{MN}{}^{PQ}
=\Bigl(\frac{1-\sigma}2Y(\id\otimes t^\alpha)\Bigr)_{MN}{}^{\langle
  PQ\rangle}\;,
\end{align}
where $\sigma$ is the permutation operator and $Y$
is the tensor that appears in the expression for generalised diffeomorphisms in extended geometry,
\begin{align}
\sigma Y= -\eta^{\alpha\beta} t_\alpha \otimes t_\beta + (\lambda,\lambda)-1+\sigma\;.
\end{align}

The existence of the THA shows that there is always
a solution to eq. \eqref{PhiLRelation2}.
The difficulty with directly analysing this equation lies in the translation
between the projections on irreducible modules in
$\adj\otimes R(\lambda)$ of the types $P_{\alpha M}{}^{\beta N}$ and
$P^\beta{}_{M,\alpha}{}^N$‚ used to characterise $\ell$ and $\varphi$,
respectively. We are not aware of any explicit translation table
in the general case, although an analysis of
the eigenvalues in eq. \eqref{egenvarde}
and the corresponding ones for $p=1$ may provide an answer.

Let us do a counting, which shows that the matrix $Q$ must be
degenerate.
Assume that $\lambda$ is a fundamental weight
(the statements may hold in a wider setting).
All irreducible modules in $\adj\otimes R(\lambda)$ appear
with multiplicity $1$. There is a single module at each level
$-(\lambda,\theta)\leq\ell\leq(\lambda,\theta)$
in the grading with respect to $\lambda$, except at $\ell=-1$ where
there are none, and at level $0$, where there is $R(\lambda)$ and in
addition a number of modules $R(\lambda+\gamma_0)$. The number
of highest roots at level $0$ equals the number of disjoint components
of the Dynkin diagram of $\fg$ when the root dual to $\lambda$ is
deleted.
The modules not in $\Phi$ are $R(\lambda)$ and
$R(\lambda+\gamma_\ell)$ for $\ell\geq1$. Their total number is
$(\lambda,\theta)+1$. The irreducible modules in $\ell$ are
$R(\lambda+\gamma_\ell)$ for $\ell\leq-2$, giving a  total
number $(\lambda,\theta)-1$.
An equation like \eqref{PhiLRelation2} would, for a generic $Q$, be
over-determined by $2$ equations. In order for a solution to exist, $Q$
must show some degeneracy, which in general will involve projections
of the two types. 
Namely, a linear combination of $P^{R(\lambda+\gamma_\ell)}{}_{\alpha
  M}{}^{\beta N}$ for $\ell\leq-2$ must have a decomposition in terms of
$P^{R(\lambda+\gamma'_{\ell'})}{}^\beta{}_{M,\alpha}{}^N$, where the
coefficients for the terms with $\gamma'=0$ and $\ell'\leq1$ agree
with those of $Q$.

The existence of the tensor hierarchy algebra thus relies on, and implies, a
quite non-trivial algebraic identity involving representation matrices
for arbitrary highest weight
representations of finite-dimensional simply laced Lie
algebras, which we have not been able to prove in an alternative way.
In Section \ref{ExamplesSection}, this identity is verified for a
number of examples, and classes of examples.
To this end, we need the eigenvalues of $Q$ when it acts on
irreducible modules in the tensor product $\adj\otimes R(\lambda)$.
They can be calculated in either picture.
We choose the $\varphi$ picture (simply because $\Phi$ contains a larger
number of irreducible modules than $L$).

The first term, $\delta_\alpha^\beta\delta_M^N$, has eigenvalue $1$ on
all modules.
The second term in $Q$, $-f_\alpha{}^{\beta\gamma}t_{\gamma M}{}^N$,
has eigenvalues that can be calculated using the quadratic Casimir operator.
We have, for any representation $R(\Lambda)\ni v$, the eigenvalue
\begin{align}
  C_2(\Lambda)v=\frac12T^\alpha\cdot T_\alpha\cdot v
  =\frac12(\Lambda,\Lambda+2\varrho)v\;.
\end{align}
For $v^\alpha{}_M$ in $R(\lambda+\gamma)$, this gives the eigenvalue
of the second term as
\begin{align}
  -C_2(\lambda+\gamma)+C_2(\lambda)+g^\vee
  =-(\lambda+\varrho,\gamma)+g^\vee+1+\delta_{\gamma,0}\;.
\end{align}
The last term, $-\frac1{(\lambda,\lambda)}(t^\beta  t_\alpha)_M{}^N$,
has eigenvalue $-\frac{2C_2(\lambda)}{(\lambda,\lambda)}$ on the
$R(\lambda)$ which is not in $\Phi$ and $0$ on the rest (including
other $R(\lambda)$'s, if $\lambda$ is not a multiple of a fundamental
weight).
The total eigenvalue of $Q$ on the module $R(\lambda+\gamma)$
becomes
\begin{align}
  Q|_{R(\lambda+\gamma)}=g^\vee-(\lambda+\varrho,\gamma)+\delta_{\gamma,0}
  -\frac{2C_2(\lambda)}{(\lambda,\lambda)}\varepsilon\;,
  \label{QValues}
\end{align}
where $\varepsilon=1$ on the $R(\lambda)$ not in $\Phi$ and $0$ otherwise.

\subsection{Comparison between $\BB$, $W$ and $S$
  at positive levels\label{IdealKSubSec}}

Consider the level decompositions of $\BB$, $W$ and $S$
in the $\mathbb{Z}$-grading
where the levels $n=p-q$ form $\fg^+$-modules (the red lines in Table
\ref{GeneralTable}).  
The modification, described in Section \ref{WSubSec}, taking us from
$\BB$ to $W$, only involves the addition of the
odd generators $f_{0i}$ at level $-1$. The
generator at $e_0$ at level $1$, remains. The generator $f_{00}$ in
$W$ is identified with $f_0$ in $\BB$ and $S$ is
obtained from $W$ by removing the generator $f_{00}$.

Since the modification only involves generators at level $-1$, it
would seem that the subalgebras containing the positive levels, which
we denote $\BB_+$, $W_+$ and $S_+$, are
unaffected, and all isomorphic. 
There are however two subtleties.

First, \apriori, there might be elements in $W$ or $S$ formed as multibrackets
with $M$ generators $e_0$ and $N$ generators $f_{0a}$ for $M \geq N$ where it is not possible to cancel the $N$ generators $f_{0a}$
against $N$ of the generators $e_0$. {\it A posteriori}, this turns out to not happen in the present case, where $\fg$ is finite-dimensional.
It follows that $\BB_+\simeq W_+$.

Second, the removal of $f_{00}$ in
the construction of $S$ may lead to the appearance of new
ideals at positive levels. Suppose there is a $\fg^+$-module
$\mu \subset W$
at some definite positive level
$n$ which does not vanish using only $[e_0,e_0]=0$, and which
furthermore obeys $[f_{0i},\mu]=0$ but
$[f_{00},\mu]\neq0$. Then $\mu$, seen as a subspace of $\wt S$, will
generate an ideal, that according to our definitions has to be factored
out to obtain the (simple) superalgebra $S$.
The positive subalgebras are isomorphic,
$S_+\simeq\BB_+$, only if there is no such ideal, and
in general $S_+=\BB_+/K$, where $K$ is the maximal ideal
of this kind.

We have no general recipe for determining whether or not the ideal $K$ of $\wt S(\fg^+)$
is non-trivial, but it is straightforward to find examples where this
is the case. Take for example $\fg^+=E_8$, and the fermionic node
attached to the fundamental (adjoint) node. The level expansion of
$\BB_+(E_8)$ (see
refs. \cite{Berman:2012vc,Cederwall:2015oua,Howe:2015hpa}) is 
\begin{align}
  \BB_+(E_8)={\bf248}_1\oplus({\bf1}\oplus{\bf3875})_2
  \oplus({\bf248}\oplus{\bf3875}\oplus{\bf147250})_3\oplus\cdots
\end{align}
where the subscripts denote the level $n$.
The elements at level $-1$ in $\BB(E_8)$ consist of the module
${\bf248}$, while $W(E_8)$ contains ${\bf248}\oplus{\bf3875}$ and
$S(E_8)$ only ${\bf3875}$ at level $-1$.
It is then obvious, just by considering tensor products of $E_8$
representations, that the singlet ${\bf1}_2\in\BB(E_8)$ generates
an ideal in $S(E_8)$, to be factored out.
A similar example occurs for $S(E_6)$. There
\cite{Berman:2012vc,Cederwall:2015oua,Howe:2015hpa},
\begin{align}
  \BB_+(E_6)&={\bf27}_1\oplus{\overline{\bf27}}_2
  \oplus{\bf78}_3\oplus\overline{\bf351}_4
  \oplus(\overline{\bf27}\oplus\overline{\bf1728})_5\nn\\
  &\quad\,\oplus({\bf1}\oplus{\bf78}\oplus{\bf650}\oplus{\bf2430}
              \oplus\overline{\bf5824})_6
  \oplus\cdots
\end{align}
At level $-1$, $S(E_6)$ contains ${\bf351}$, but not the
$\overline{\bf27}$ present in $\BB(E_6)$ and $W(E_6)$. The singlet
at level $6$ generates an ideal.



\section{The embeddings $\wt W(\fg) \subset \wt S(\fg^+) \subset \wt W(\fg^+)$
\label{EmbeddingSection}}

Suppose that $\lambda$ is a fundamental weight, which we take to be $\Lambda_2$ for simplicity. Thus node $1$ is connected to node $2$
with a single line but disconnected from nodes $3,4,\ldots,r+1$. We will here show that in this case $\wt S(\fg)$ and $\wt W(\fg)$ can be embedded in
$\wt S(\fg^+)$ as subalgebras at height $q=0$ with node $2$ in $\wt S(\fg^+)$ as ``node $1$'' in $\wt S(\fg)$ and $\wt W(\fg)$.
First we set
\begin{align}
e_0{}' &=[e_0,e_1]\;, & f_{0i}{}'&=-[f_{0i},f_1]\;, & e_j{}'&=e_j\;, &f_j{}'&=f_j
\end{align}
for $i=3,\ldots,r+1$ for $j=2,3,\ldots,r+1$.
This already gives an embedding of $\wt S(\fg)$ in $\wt S(\fg^+)$. In order to extend it $\wt S(\fg)$ to $\wt W(\fg)$, we have
to find elements $f_{00}{}'$ and $h_0{}'$ in $\wt S(\fg^+)$. They will have the form $f_{00}{}'=f_{0\alpha}$ and $h_0{}'=h_\mu$
for some $\alpha,\mu \in \mathfrak{h}^\ast$, where $\mu$ must satisfy
\begin{align} \label{muegenskaper}
(\mu,\alpha_0+\alpha_1)&=0\;,& (\mu,\alpha_2)&=-1\;, & (\mu,\alpha_3)&=\cdots=(\mu,\alpha_{r+1})=0\;.
\end{align}
From the relation $[e_0{}',f_{00}{}']=h_{0}{}'$ we then get
\begin{align} \label{alfamurelation}
\alpha-(\alpha,\alpha_1)\alpha_1 = \mu\;.
\end{align}
If we now set (recall that we assume $(\lambda,\lambda)\neq 1$)
\begin{align}
\mu&=\frac{\lambda+(\lambda,\lambda)\alpha_1}{(\lambda,\lambda)-1}\;,&
\alpha&=\frac{\lambda}{(\lambda,\lambda)-1}\;,
\label{MuAlphaEq}
\end{align}
then it is easy to show that these element satisfy the conditions (\ref{muegenskaper}) and (\ref{alfamurelation}),
and then the defining relations for $\wt W(\fg)$ follow. Thus $\wt W(\fg) \subset \wt S(\fg^+)$.
Since clearly also $\wt S(\fg) \subset \wt W(\fg)$ and $\wt S(\fg^+) \subset \wt W(\fg^+)$ we have a chain of embeddings
\begin{align} \label{inbaddningar}
\wt W(\fg^+) \supset  \wt S(\fg^+) \supset \wt W(\fg) \supset \wt S(\fg) 
\end{align}
that can be continued to lower rank at least as long as the grey node is connected to only one white node, so that chain of embedding corresponds
to a chain of white nodes, 
but presumably our definition of the tensor hierarchy algebras can be generalised in order to allow for more than one ``node 1''
so that the chain could be continued in general (and of course also to higher rank with the definitions that we have 
already). The procedure is similar to the one giving rise to a
chain of embeddings 
for the corresponding Borcherds superalgebras, described in ref. \cite{Kleinschmidt:2013em}.

To what extent do the embeddings (\ref{inbaddningar}) hold if we
``remove the tildes'', \ie, if we factor out the 
maximal ideal intersecting the subalgebra at $p=0$ trivially?
Ideals at negative levels will not affect the subalgebra embeddings,
since level $1$ is identical in $S$ and $W$. We need to investigate
what happens when there is ideal $K\subset\wt S$ (see Section
\ref{IdealKSubSec}) at positive levels
which is not an ideal in $\wt W$.
Then there is not a subalgebra embedding $S(\fg^+)\subset
W(\fg^+)$. If the ideal $K$ is non-trivial, one instead has
\begin{align}
  S(\fg^+)\ltimes K\subset W(\fg^+)\;.
\end{align}
We already know that $W_+(\fg)\simeq\BB_+(\fg)$ (see Section
\ref{IdealKSubSec}).
The only ideal factored out at positive
levels to arrive at the simple superalgebra $\BB(\fg)$ is the one
generated by $[e_0,e_0]$ \cite{Cederwall:2015oua}. This implies that
$W_+(\fg)=\wt W_+(\fg)$, so the ideal $K$ in $\wt S(\fg^+)$ intersects
$\wt W(\fg)$ trivially. 
We thus have a subalgebra embedding
\begin{align}
  W(\fg) \subset S(\fg^+)
\end{align}
This can be observed in the examples of Section \ref{IdealKSubSec}.
In both examples, the singlet generating the ideal appears at $q=2$, and
the ideal does not intersect $q=0$ (the locus of the $W(\fg)$ subalgebra).

\section{Examples\label{ExamplesSection}}

In this Section, we give a number of examples of tensor hierarchy
algebras. Focus is put on the identity \eqref{PhiLRelation}, which is
the crucial test for the existence of the algebras. Even if it follows
from the construction that the Jacobi identities are satisfied, the
proof is quite implicit. Therefore, we want to verify it explicitly in
some concrete cases. We give them by increasing value of
$(\lambda,\theta)$ (and subsequently, increasing degree of complication),
from $1$ to $3$.

\subsection{$(\lambda,\theta)=1$}

Consider the situation when $(\lambda,\theta)=1$, \ie, when
$\tR_1=\nullrep$. Then, $\ell_{\alpha M}{}^{\beta N}=0$ and $\varphi=Q$.
The invariant tensor $\varphi$ will have vanishing projections on
$R(\lambda+\theta)$ and $R(\lambda)$. We calculate the eigenvalues on
these modules using eq. \ref{QValues}, and get
\begin{align}
  \varphi|_{R(\lambda+\theta)}&=g^\vee-1-(\varrho,\theta)=0\;,\nn\\
  \varphi|_{R(\lambda)}&=g^\vee+1-\frac{2C_2(\lambda)}{(\lambda,\lambda)}\;.
\end{align}

The vanishing of the latter expression can be shown as follows.
The condition $(\lambda,\theta)=1$ means that $\lambda$ must be a fundamental weight
$\Lambda_i$ corresponding to a simple root $\alpha_i$ (and furthermore that the associated Coxeter label is $1$).
Let $\fg^{-}$ be the simple subalgebra of $\fg$ with Dynkin diagram obtained by removing node $i$ from the diagram of $\fg$.
The grading of $\fg$ with respect to $\lambda$ is a $3$-grading: 
\begin{align}
\fg = \fg_{(-1)} \oplus \fg_{(0)} \oplus \fg_{(1)} = \fg_{(-1)} \oplus (\fg^- \oplus \mathbb{R}) \oplus \fg_{(1)}\;,
\end{align}
where $\fg_{(1)}$ is a module for a $\fg^-$ representation $R(-\nu)=R(\overline \nu)$
with $e_i$ as a lowest weight state and $e_\theta$ as a highest weight state.
Thus $\theta - \alpha_i = \overline \nu - (-\nu) $. However, $\theta$ and $\alpha$ are roots of $\fg$, whereas
$\overline \nu$ and $\nu$ are weights of $\fg^-$ and thus linear combinations of only the
simple roots $\alpha_j$ such that $j \neq i$.
For $\nu$ we can determine this linear combination by the conditions
$(\nu,\lambda)=0$, which means that $\nu$ has zero coefficient for $\alpha_i$ in the basis of simple roots, and $(\nu,\alpha_j)=-(\alpha_i,\alpha_j)$ if $i\neq j$.
We then get
\begin{align}\label{andra}
\nu = \frac{\lambda}{(\lambda,\lambda)}-\alpha_i
\end{align}
which gives
\begin{align}
(\varrho,\nu)&=\frac{(\varrho,\lambda)}
{(\lambda,\lambda)}-1\;.
\end{align}
Now $\overline{\nu}$ is the image of $\nu$ under an isometry of the weight lattice that permutes the simple roots of $\fg^{-}$ (which is just the identity map,
$\overline \nu = \nu$, unless the symmetry group of the Dynkin diagram of $\fg^-$ is $\mathbb{Z}_2$), and since the Weyl vector $\varrho$ of $\fg$
has the property $(\varrho,\alpha_j)=1$ for all simple roots $\alpha_j$ of $\fg$ (in particular those of $\fg^-$), we have $(\varrho,\overline\nu)=(\varrho,\nu)$.
Taking the inner product of $\varrho$ with $\theta=\alpha_i+\nu+\overline{\nu}$ we then get
\begin{align}
(\varrho,\theta)&=(\varrho,\alpha_i)+2(\varrho,\nu)
=1+2\frac{(\varrho,\lambda)}{(\lambda,\lambda)}-2(\varrho,\alpha_i)
=2\frac{(\varrho,\lambda)}{(\lambda,\lambda)}-1\;.
\end{align}
Using the expression for the second Casimir of a representation
$R(\Lambda)$ with highest weight $\Lambda$,
$C_2(\Lambda)=\frac12(\Lambda,\Lambda+2\varrho)$, this relation  may be
expressed as
\begin{align}
\frac{2C_2(\lambda)}{(\lambda,\lambda)}=g^\vee+1\;,
\end{align}
or equivalently, using the Freudenthal--de Vries ``strange formula'',
\begin{align}
\frac{6(\lambda,\lambda)(\varrho,\varrho)}{(\lambda,\varrho)}=\dim\fg\;.
\end{align}

\subsection{$(\lambda,\theta)=2$: the THA over an affine algebra} \label{affinesection}

We consider the case when $R(\lambda)$ is the adjoint of $\fg$ so that
$\fg^+$ is the affine extension of $\fg$ 
(for example $\fg=E_8$ and $\fg^+=E_9$).
We thus take $\lambda=\theta$, \ie, $(\lambda,\theta)=2$.

The representations in $\adj\otimes\adj$ which are
not in $\Phi$ are $R(2\theta)$,
$\bigoplus_iR(2\theta-\alpha_i)$ and $R(\theta)$, at levels $2$, $1$ and $0$
respectively, where $\{\alpha_i\}$ is the set of simple roots with
$(\theta,\alpha_i)=1$.
The eigenvalues of $Q$ on these representations are
$-1$, $1$ and $1$, respectively.

The modules in $\Phi$ are a number of $R(\theta+\gamma_0)$,
where $\gamma_0$ are the highest roots at level $0$, and
$R(0)={\bf1}$. Each $\gamma_0$ defines a subalgebra $\fg_{\gamma_0}$,
the Dynkin diagram of which is a component of 
the Dynkin diagram of $\fg$ with the node(s) corresponding to $\theta$
removed, and $\lambda+\gamma_0$ is the highest root of $\fg_{\gamma_0}$. 

For the example $\fg=E_8$‚ $\fg^+=E_9$, we have $\adj={\bf248}$. There
is a single root $\gamma_0$, and
${\bf248}\otimes{\bf248}={\bf27000}\oplus{\bf3875}\oplus{\bf1}
\oplus{\bf30380}\oplus{\bf248}$. Of these,
${\bf3875}\oplus{\bf1}$ are contained in $\Phi$.

Tables \ref{AffineWTableBasis} and \ref{AffineSTableBasis} show the
local ($p=-1,0,1$) parts of $W(\fg^+)$ and $S(\fg^+)$.
Tables \ref{E9WTable} and \ref{E9STable} give the corresponding
decompositions of $W(E_9)$ and $S(E_9)$ into $E_8$ modules.

\begin{table}
  \begin{align*}
  \xymatrix@=.4cm{
    \ar@{-}[]+<2em,1em>;[dddd]+<2em,-1em>
    \ar@{-}[]+<-0.8cm,-1em>;[rrr]+<1.4cm,-1em>
&\ar@{-}[]+<6em,1em>;[dddd]+<6em,-1em> p=-1 & \ar@{-}[]+<4em,1em>;[dddd]+<4em,-1em> p=0 &p=1\\
q=2&&&L^\sh\\
q=1&\varphi^\sh\quad\quad\Phi_{\alpha\beta}^\sh\quad\quad G_\alpha^\sh&{f_{00}}\quad\quad T_\alpha^\sh& E_\alpha^\sh\quad\quad{L}
       \\ 
q=0&{F_\alpha}\quad\quad{\varphi}\quad\quad\Phi_{\alpha\beta}\quad\quad G_\alpha
&{k}\quad\quad{\tilde k}\quad\quad{T_\alpha}
           & E_\alpha\\ 
q=-1 & F_\alpha^\fl & {e_0} 
  }
\end{align*}
  \caption{\it Basis elements of $W$ when $\fg^+$ is the affine
    extension of $\fg$.}
\label{AffineWTableBasis}
\end{table}

\begin{table}
  \begin{align*}
  \xymatrix@=.4cm{
    \ar@{-}[]+<2em,1em>;[dddd]+<2em,-1em>
    \ar@{-}[]+<-0.8cm,-1em>;[rrr]+<1.4cm,-1em>
    &\ar@{-}[]+<4.5em,1em>;[dddd]+<4.5em,-1em> p=-1
    & \ar@{-}[]+<2.9em,1em>;[dddd]+<2.9em,-1em> p=0 &p=1\\
q=2&&&L^\sh\\
q=1&\varphi^\sh\quad\quad\Phi_{\alpha\beta}^\sh&T_\alpha^\sh& E_\alpha^\sh\quad\quad{L}
       \\ 
q=0&{H_\alpha}\quad\quad{\varphi}\quad\quad\Phi_{\alpha\beta}
&{\tilde k}\quad\quad{T_\alpha}
           & E_\alpha\\ 
q=-1 & H_\alpha^\fl & {e_0} 
  }
\end{align*}
  \caption{\it Basis elements of $S$ when $\fg^+$ is the affine
    extension of $\fg$.}
\label{AffineSTableBasis}
\end{table}

\begin{table}
  \begin{align*}
  \xymatrix@=.4cm{
    \ar@{-}[]+<1.1cm,1em>;[dddd]+<1.1cm,-1em>
    \ar@{-}[]+<-0.8cm,-1em>;[rrrr]+<1.4cm,-1em>
&p=-1 &p=0 &p=1&p=2\\
q=2&&&\Blue{\bf1}&{\bf248}\\
q=1&\Blue{\bf1}\oplus\Blue{\bf3875}\oplus\Blue{\bf248}&{\bf 1}\oplus\Blue{\bf248}& {\bf248}\oplus\Blue{\bf1}
       &{\bf1}\oplus{\bf3875}\oplus{\bf248}\\
q=0&{\bf248}\oplus\Blue{\bf1}\oplus\Blue{\bf3875}\oplus\Blue{\bf248}
&{\bf 1}\oplus{\bf248}\oplus{\bf 1}
           & {\bf248}&{\bf1}\oplus{\bf3875}\\
q=-1 & {\bf248} & {\bf 1} 
  }
\end{align*}
  \caption{\it Basis elements of $W(E_9)$. The modules not present in ${\scr B}(E_9)$ are marked
  blue. Note the presence of $\tR_1={\bf1}$.}
\label{E9WTable}
\end{table}

\begin{table}
  \begin{align*}
  \xymatrix@=.4cm{
    \ar@{-}[]+<1.1cm,1em>;[dddd]+<1.1cm,-1em>
    \ar@{-}[]+<-0.8cm,-1em>;[rrrr]+<1.4cm,-1em>
&p=-1 &p=0 &p=1&p=2\\
q=2&&&{\bf1}&{\bf248}\\
q=1&{\bf1}\oplus{\bf3875}&{\bf248}& {\bf248}\oplus{\bf1}
       &{\bf1}\oplus{\bf3875}\oplus{\bf248}\\
q=0&{\bf248}\oplus{\bf1}\oplus{\bf3875}
&{\bf 1}\oplus{\bf248}
           & {\bf248}&{\bf1}\oplus{\bf3875}\\
q=-1 & {\bf248} & {\bf 1} 
  }
\end{align*}
  \caption{\it Basis elements of $S(E_9)$. Note the symmetry under
  $(p,q)\leftrightarrow(1-p,1-q)$ associated with existence of a
    bilinear form.}
\label{E9STable}
\end{table}

The
eigenvalue of $Q$ on $R(\theta+\gamma_0)$ is
$g^\vee-g_{\gamma_0}^\vee+1$. The eigenvalue on ${\bf1}$ is $2g^\vee+1$.
The projector on ${\bf1}$ in the $\ell$ picture is 
$\frac1{\dim\fg}\eta_{\alpha M}\eta^{\beta N}$, and its eigenvalues on the
modules in the $\varphi$ picture are $\pm\frac{1}{\dim\fg}$,
depending on whether
it is in the symmetric or antisymmetric part of the tensor product. We
saw that $Q$ has eigenvalue $-1$ on the symmetric module not in
$\Phi$, and $1$ on the antisymmetric ones.

Equation \eqref{PhiLRelation2} is solved with
\begin{align}
\varphi&=\sum\limits_{\gamma_0\in
  H_0}(g^\vee-g_{\gamma_0}^\vee+2)P_{R(\theta+\gamma_0)}
         +2(g^\vee+1)P_{\bf1}\;,\nn\\
\ell&=\dim\fg\, P_{\bf1}\;,
\end{align}
where the projectors in
$\varphi$ and $\ell$ are expressed in their respective bases.
In the example with $S(E_9)$, $g^\vee=30$ and $\fg_{\gamma_0}=E_7$ with
$g^\vee_{\gamma_0}=18$, and we get
$\varphi=14P_{\bf3875}+62P_{\bf1}$, \ie\ (see eq. \eqref{EEightProjs}),
\begin{align}
  \varphi_{\alpha\beta}{}^{\gamma\delta}=
2\delta_{(\alpha}^\gamma\delta_{\beta)}^\delta
-f_{(\alpha}{}^{\gamma\epsilon}f_{\beta)}{}^\delta{}_\epsilon\;.
\end{align}
This
  latter expression is generic in the present class of examples. This
  can be shown by
  inserting this expression for $\varphi$, together with
  $\ell_{\alpha\beta}{}^{\gamma\delta}=\eta_{\alpha\beta}\eta^{\gamma\delta}$,
into eq. \eqref{PhiLRelation}
  with $(t_\alpha)_\beta{}^\gamma=-f_{\alpha\beta}{}^\gamma$ and using
the Jacobi identity. 

In this series of examples, $\fg^+$ is the (untwisted) affine algebra over
$\fg$. At level $1$, there is an anti-fundamental module,
whose lowest weight state is $e_0$. At level $0$, there is, in addition
to the adjoint, a single generator $L$, which can be identified with
the Virasoro generator $L_1$. At level $-1$, we find a shifted
fundamental module, with highest weight state $L^\sh$.

As can be seen in Table \ref{E9WTable}, there is a symmetry in the
representation content of $S(E_9)$ under 
$(p,q)\leftrightarrow(1-p,1-q)$, associated with the existence of an invariant
non-degenerate bilinear form \cite{Palmkvist:2013vya,Bossard:2017wxl}. This symmetry occurs for $S(\fg^+)$ whenever $\fg$ is an affine algebra.
In general, if there is an affine Kac--Moody algebra $\fg^{(k)}$ obtained by adding a chain of $k$ white nodes to the Dynkin diagram of $\fg$
(for example if $\fg=E_{9-k}$),
then there is such a symmetry under $(p,q)\leftrightarrow(k-p,1-q)$ in $S(\fg)$ \cite{Palmkvist:2013vya}, and this seems to hold even for negative $k$
(if ``adding a chain of $k$ white nodes'' is interpreted as ``removing a chain of $-k$ white nodes'') \cite{Bossard:2017wxl}.

\subsection{$(\lambda,\theta)=2$: another series}

Another series of examples, also with $(\lambda,\theta)=2$, is $D_r$ with
$R(\lambda)$ a $3$-form, $(0010\ldots0)$.
Then, $\lambda+\theta=(0110\ldots0)$,
$\lambda+\gamma_1=(10010\ldots0)$ and
$\lambda-\beta_2=(10\ldots0)$.
The modules that are not part of $\Phi$ are
$R(\lambda+\theta)$, $R(\lambda+\gamma_1)$ and $R(\lambda)$.
The eigenvalues of $Q$ on them are $-1$, $2$ and $2$, respectively.
The projector on $R(\lambda-\beta_2)$ is proportional to
$\delta^{\mathstrut}_{mn,[pq}\delta_{\mathstrut}^{st,[uv}\delta^{w]}_{r]}$. Letting
this tensor act on states $\Psi_{st}{}^{pqr}$ in the three
modules that do not appear in $\Phi$, one finds the eigenvalues
$\frac13$, $-\frac23$ and $-\frac23$, which with a factor $3$ cancels
the contribution from $Q$,  and eq. \eqref{PhiLRelation2} holds.
The extended algebra $\fg^+$ is hyperbolic for $r\leq9$.

\subsection{An example with $(\lambda,\theta)=3$}

Finally, we would like to give an example where
$(\lambda,\theta)=3$. With $\fg=E_8$ and
$\adj=\rep{248}=\Weight10000000$, we take
$\lambda=\Weight01000000$, $R(\lambda)=\rep{30\,380}$.

The construction makes use of the projections on the irreducible
representations in $\adj\otimes\adj$, which are
\begin{align}
  P^\rep{27\,000}{}_{\alpha\beta}{}^{\gamma\delta}
  &=\frac67\delta_{(\alpha}^\gamma\delta_{\beta)}^\delta
  +\frac1{14}f_{(\alpha}{}^{\gamma\epsilon}f_{\beta)}{}^\delta{}_\epsilon
  +\frac3{217}\eta_{\alpha\beta}\eta^{\gamma\delta}\;,\nn\\
  P^\rep{30\,380}{}_{\alpha\beta}{}^{\gamma\delta}&=\delta_{\alpha\beta}^{\gamma\delta}
  +\frac1{60}f_{\alpha\beta}{}^\epsilon f^{\gamma\delta}{}_\epsilon\;,\nn\\
  P^\rep{3\,875}{}_{\alpha\beta}{}^{\gamma\delta}
  &=\frac17\delta_{(\alpha}^\gamma\delta_{\beta)}^\delta
  -\frac1{14}f_{(\alpha}{}^{\gamma\epsilon}f_{\beta)}{}^\delta{}_\epsilon
  -\frac1{56}\eta_{\alpha\beta}\eta^{\gamma\delta}\;,\label{EEightProjs}\nn\\
  P^\rep{248}{}_{\alpha\beta}{}^{\gamma\delta}
  &=-\frac1{60}f_{\alpha\beta}{}^\epsilon f^{\gamma\delta}{}_\epsilon\;,\nn\\
  P^\rep1{}_{\alpha\beta}{}^{\gamma\delta}
  &=\frac1{248}\eta_{\alpha\beta}\eta^{\gamma\delta}\;.
\end{align}
The only identity, not following from the Jacobi identities, that is
needed for verification of the projector properties is
\begin{align}
  q_{\alpha\beta}{}^{\kappa\lambda}q_{\kappa\lambda}{}^{\gamma\delta}
  =24\delta_{(\alpha}^\gamma\delta_{\beta)}^\delta-10q_{\alpha\beta}{}^{\gamma\delta}
  +12\eta_{\alpha\beta}\eta^{\gamma\delta}\;,
\end{align}
where
$q_{\alpha\beta}{}^{\gamma\delta}=f_{(\alpha}{}^{\gamma\epsilon}f_{\beta)}{}^\delta{}_\epsilon$. 
Define
${\star}P^R{}_{\alpha\beta}{}^{\gamma\delta}=\eta_{\alpha\epsilon}\eta^{\gamma\varphi}
P^R{}_{\varphi\beta}{}^{\epsilon\delta}$. Then,
${\star}P^{R_i}=\sum\limits_{j=1}^5M_{ij}P^{R_j}$, where
$\{R_i,i=1,\ldots,5\}$ are the representations in the order listed
above, and
\begin{align}
  M=\left(
  \begin{matrix}
    \frac{23}{62}&\frac{90}{217}&\frac{27}{31}&\frac{225}{62}&\frac{3375}{31}\\
    \frac7{15}&\frac12&\frac7{10}&0&-\frac{245}2\\
    \frac18&\frac5{56}&-\frac38&-\frac{25}8&\frac{125}8\\
    \frac1{30}&0&-\frac15&\frac12&-1\\
    \frac1{248}&-\frac1{248}&\frac1{248}&-\frac1{248}&\frac1{248}
  \end{matrix}
  \right)
\end{align}
This translation matrix is used in some of the following calculations.

\setlength\arraycolsep{1.4pt}
The representations in $\adj\otimes R(\lambda)$
obtained from the roots in eq. \eqref{RlambdaAdjRepr} are:
\begin{align}
\begin{array}{rccclrclrcl}
  \theta&=&\gamma_3&=&\Root23456423\phantom{xxxxx}
  &  \lambda+\gamma_3&=&\Weight11000000\phantom{xxxxx}
  &  R(\lambda+\gamma_3)&=&\rep{4\,096\,000} \nn\\
 && \gamma_2&=&\Root12456423 & \lambda+\gamma_2&=&\Weight00100000
  &  R(\lambda+\gamma_2)&=&\rep{2\,450\,240} \nn\\
  &&\gamma_1&=&\Root11234322 & \lambda+\gamma_2&=&\Weight10000010
  &  R(\lambda+\gamma_1)&=&\rep{779\,247} \nn\\
  &&\gamma_0&=&\Root00123212 & \lambda+\gamma_0&=&\Weight00000001
  &  R(\lambda+\gamma_0)&=&\rep{147\,250} \nn\\
  &&\gamma'_0&=&\Root10000000 & \lambda+\gamma'_0&=&\Weight20000000
  &  R(\lambda+\gamma'_0)&=&\rep{27\,000} \\
                          &&&&& \lambda&=&\Weight01000000
  &  R(\lambda)&=&\rep{30\,380} \nn\\
  &&\beta_2&=&\Root12222101 & \lambda-\beta_2&=&\Weight00000010
  &  R(\lambda-\beta_2)&=&\rep{3\,875} \nn\\
  &&\beta_3&=&\Root13456423 & \lambda-\beta_3&=&\Weight10000000
  &  R(\lambda-\beta_3)&=&\rep{248}
\end{array}
\end{align}
(we use the notation $\Root{i_1}{i_2}{i_3}{i_4}{i_5}{i_6}{i_7}{i_8}$
for coefficients in root basis and
$\Weight{j_1}{j_2}{j_3}{j_4}{j_5}{j_6}{j_7}{j_8}$ in weight basis).

To distinguish the projectors from the ones for $\adj\otimes\adj$, we
denote them
$\PP^R{}_{\alpha,\beta\gamma}{}^{\delta,\epsilon\varphi}$. They satisfy
$P^\rep{248}{}_{\beta\gamma}{}^{\kappa\lambda}
\PP^R{}_{\alpha,\kappa\lambda}{}^{\delta,\epsilon\varphi}=0$, which can
be implemented by letting
\begin{align}
  \PP^R{}_{\alpha,\beta\gamma}{}^{\delta,\epsilon\varphi}
  =P^\rep{30\,380}{}_{\beta\gamma}{}^{\kappa\lambda}
  \Pi^R{}_{\alpha,\kappa\lambda}{}^{\delta,\rho\sigma}
  P^\rep{30\,380}{}_{\rho\sigma}{}^{\epsilon\varphi}\;.
\end{align}
The $\Pi$'s are equivalent modulo combinations of an  antisymmetric
pair into $\rep{248}$‚ which we will treat as equality.
The relevant product and trace on the $\Pi$'s are
\begin{align}
  (\Pi\circ\Pi')_{\alpha,\beta\gamma}{}^{\delta,\epsilon\varphi}
  &=\Pi_{\alpha,\beta\gamma}{}^{\kappa,\lambda\mu}
  P^\rep{30\,380}{}_{\lambda\mu}{}^{\rho\sigma}
  \Pi'_{\kappa,\rho\sigma}{}^{\delta,\epsilon\varphi}\;,\nn\\
  \tr\Pi&=\Pi_{\alpha,\beta\gamma}{}^{\alpha,\epsilon\varphi}
  P^\rep{30\,380}{}_{\epsilon\varphi}{}^{\beta\gamma}\;.
\end{align}
The explicit forms of the $\Pi^R$'s are
\begin{align}
\Pi^\rep{4\,096\,000}{}_{\alpha,\beta\gamma}{}^{\delta,\epsilon\varphi}
&=\frac43\delta_{(\alpha}^\delta\delta_{\beta)\gamma}^{\epsilon\varphi}
-\bigl(\Pi^\rep{779\,247}
+\Pi^\rep{147\,250}
+\frac{14}{45}\Pi^\rep{27\,000}\bigr.\nn\\
&\qquad\qquad\qquad+\bigl.\frac13\Pi^\rep{30\,380}
+\frac7{15}\Pi^\rep{3\,875}
+\Pi^\rep{248}\bigr){}_{\alpha,\beta\gamma}{}^{\delta,\epsilon\varphi}
\;,\nn\\
\Pi^\rep{2\,045\,240}{}_{\alpha,\beta\gamma}{}^{\delta,\epsilon\varphi}
&=\delta_{\alpha\beta\gamma}^{\delta\epsilon\varphi}
-\bigl(\frac{31}{45}\Pi^\rep{27\,000}
+\frac23\Pi^\rep{30\,380}
+\frac8{15}\Pi^\rep{3\,875}\bigr){}_{\alpha,\beta\gamma}{}^{\delta,\epsilon\varphi}\;,\nn\\
\Pi^\rep{779\,247}{}_{\alpha,\beta\gamma}{}^{\delta,\epsilon\varphi}
&=\frac{49}{26}(U+2V){}_{\alpha,\beta\gamma}{}^{\delta,\epsilon\varphi}\;,\nn\\
\Pi^\rep{147\,250}{}_{\alpha,\beta\gamma}{}^{\delta,\epsilon\varphi}
&=\frac16(U-14V){}_{\alpha,\beta\gamma}{}^{\delta,\epsilon\varphi}\;,\nn\\
\Pi^\rep{27\,000}{}_{\alpha,\beta\gamma}{}^{\delta,\epsilon\varphi}
&=-\frac{15}{434}f_{\alpha\beta}{}^\rho f^{\delta\epsilon}{}_\sigma
    P^\rep{27\,000}{}_{\gamma\rho}{}^{\varphi\sigma}\;,\nn\\
\Pi^\rep{30\,380}{}_{\alpha,\beta\gamma}{}^{\delta,\epsilon\varphi}
    &=-\frac1{30}f_{\alpha\beta}{}^\rho f^{\delta\epsilon}{}_\sigma
    \delta_{\gamma\rho}^{\varphi\sigma}\;,\nn\\
\Pi^\rep{3\,875}{}_{\alpha,\beta\gamma}{}^{\delta,\epsilon\varphi}
&=-\frac5{168}f_{\alpha\beta}{}^\rho f^{\delta\epsilon}{}_\sigma
    P^\rep{3\,875}{}_{\gamma\rho}{}^{\varphi\sigma}\;,\nn\\
\Pi^\rep{248}{}_{\alpha,\beta\gamma}{}^{\delta,\epsilon\varphi}
&=\frac2{245}\eta_{\alpha\beta}\eta^{\delta\epsilon}\delta_\gamma^\varphi\;,
\end{align}
where
\begin{align}
  U_{\alpha,\beta\gamma}{}^{\delta,\epsilon\varphi}
  &=P^\rep{3\,875}{}_{\alpha\beta}{}^{\delta\epsilon}\delta_\gamma^\varphi
  -\bigl(\frac1{28}\Pi^\rep{30\,380}+\frac7{20}\Pi^\rep{3\,875}
 +\frac1{16}\Pi^\rep{248}\bigr){}_{\alpha,\beta\gamma}{}^{\delta,\epsilon\varphi}\nn\;,\\
 V_{\alpha,\beta\gamma}{}^{\delta,\epsilon\varphi}
 &=P^\rep{3\,875}{}_{\alpha\beta}{}^{\varphi\rho}
 P^\rep{3\,875}{}_{\gamma\rho}{}^{\delta\epsilon}\nn\\
 &\quad-\bigl(\frac{11}{392}\Pi^\rep{30\,380}-\frac7{40}\Pi^\rep{3\,875}
 +\frac3{128}\Pi^\rep{248}\bigr){}_{\alpha,\beta\gamma}{}^{\delta,\epsilon\varphi}\;.
\end{align}
It is relatively straightforward to show that
$U\circ U=\frac12(U-V)$. The remaining identity needed is
$U\circ V=-\frac1{56}(U-40V)$, from which it then follows that
$V\circ V=\frac1{392}(10U-127V)$.
We have not checked it explicitly, but it is needed for the
projection operators to work and to give the
correct dimensions of the representations.

We now want to translate between the two ``pictures'', \ie, express
${\star}\PP^R{}_{\alpha,\beta\gamma}{}^{\delta,\epsilon\varphi}
\equiv\eta_{\alpha\rho}\eta^{\delta\sigma}
\PP^R{}_{\sigma,\beta\gamma}{}^{\rho,\epsilon\varphi}$ in terms of the
$\PP^R$'s.
This needs to be done for $R(\lambda-\beta_3)=\rep{248}$ and
$R(\lambda-\beta_2)=\rep{3\,875}$.
A lengthy calculation yields
\begin{align}\label{StarPP}
  {\star}\PP^\rep{248}&=\frac1{245}\bigl(\PP^\rep{4\,096\,000}
  -2\PP^\rep{2\,450\,240}+\PP^\rep{779\,247}+\PP^\rep{147\,250}\nn\\
  &\qquad\qquad-\frac{16}{15}\PP^\rep{27\,000}-\PP^\rep{30\,380}
  -\frac35\PP^\rep{3\,875}+\PP^\rep{248}
  \bigr)\;,\nn\\
  {\star}\PP^\rep{3\,875}&=\frac5{784}\bigl(\PP^\rep{4\,096\,000}
  +0\PP^\rep{2\,450\,240}-\frac{23}3\PP^\rep{779\,247}+15\PP^\rep{147\,250}
  \nn\\
  &\qquad\qquad+\frac{98}5\PP^\rep{27\,000}-7\PP^\rep{30\,380}
  -\frac{49}5\PP^\rep{3\,875}-6\PP^\rep{248}
  \bigr)\;. 
\end{align}
A good check on the result is that the dimensions add up correctly.

In order for eq. \eqref{PhiLRelation} to have a solution,
\ie, for the tensor hierarchy
algebra to exist, it must be possible to cancel the contribution from
$Q$ to the representations $\rep{4\,096\,000}$, $\rep{2\,045\,240}$,
$\rep{779\,247}$ and $\rep{30\,380}$
by a  linear combination of the right hand sides of
eq. \eqref{StarPP}.
The decomposition of $Q$ is given by eq. \eqref{QValues}, and we have
\begin{align}
  Q&=-2\PP^\rep{4\,096\,000}+\PP^\rep{2\,045\,240}
  +11\PP^\rep{779\,247}+19\PP^\rep{147\,250}\nn\\
  &\quad+29\PP^\rep{27\,000}
  +11\PP^\rep{30\,380}+43\PP^\rep{3\,875}+61\PP^\rep{248}\;.
\end{align}
The coefficients of the projectors on the representation not present
in $\Phi$ cancel by adding
\begin{align}
    \ell=\frac{1176}5{\star}\PP^\rep{3\,875}+\frac{245}2{\star}\PP^\rep{248}
\end{align}
as $(-2,1,11,11)+(\frac32,0,-\frac{23}2,-\frac{21}2)
+(\frac12,-1,\frac12,-\frac12)
=0$.
The remainder is
\begin{align}
  \varphi=Q+\ell=42\PP^\rep{147\,250}+\frac{868}{15}\PP^\rep{27\,000}
  +28\PP^\rep{3\,875}+\frac{105}2\PP^\rep{248}
  \;.
\end{align}

The extended algebra $\fg^+$ in this example is the hyperbolic Lie
algebra $D_7^{++}$.
In the tensor hierarchy algebra $S$, level $1$ contains
$R(-\Lambda)$, where $\Lambda=\DPPWeight100000000$. At level $0$,
there is of course the adjoint, but also (at least) two lowest weight
representations $R(-\mu)$, $R(-\nu)$, with
$\mu=\DPPWeight000000010$, $\nu=\DPPWeight000000100$,
whose lowest representations in a grading with
respect to the extending node are the $\rep{248}$ and $\rep{3\,875}$
in $L$.

\section{Conclusions} \label{consec}

We have given definitions of the tensor hierarchy algebras $W(\fg^+)$
and $S(\fg^+)$ in terms of generators and relations, when $\fg^+$ is a
Lie algebra obtained by extending the finite-dimensional (simply
laced) Lie algebra $\fg$ by a single node.
A number of examples are given, of which some are relevant to physical
applications. 

One main difficulty with deriving the content of the superalgebras is
associated with the appearance of ``mixed'' elements; the root space
contains roots where the coefficients for the simple roots are not all
positive or all negative. This phenomenon is also associated with the
appearance of ``extra'' elements together with $\fg^+\oplus{\mathbb R}$ at
level $n=0$ (beginning with the generators $L_{\alpha M}$). This is
seen \eg\ in Tables \ref{GeneralWTableBasis} and
\ref{GeneralSTableBasis}.
Such elements are significant in the application to extended geometry,
as explained in ref. \cite{CederwallPalmkvistTHAII}.

The definition should be good also for infinite-dimensional $\fg$.
The derivations in the present paper will then not be valid. For
example, there will typically also appear some elements in $\tR_0$,
\ie, a pair of isomorphic modules
at $(p,q)=(0,1)$ and $(0,2)$ in the double
grading. We have verified this for affine $\fg$, where $\tR_0$ is a
singlet, and $\tR_p=\nullrep$, $p<0$. For ``more
infinite-dimensional'' algebras, \eg\ hyperbolic $\fg$, also
$\tR_{-1}$ etc. can appear.
Even if the definitions remain formally identical, the implications
seem to differ drastically, also in the local subalgebra.
It would be desirable to design a method that determines the ``extra''
elements in a more direct way. For infinite-dimensional $\fg$, there may also be ``extra'' elements
at positive levels $n$, so that it would no longer be true that $W_+ \simeq \scr  B_+$,
as stated for finite-dimensional $\fg$ in Section \ref{IdealKSubSec}.

A topic we have not touched is representation theory for THA's.
In particular, the construction of non-trivial representations would be
a more efficient and general method to prove that the tensor hierarchy algebra is non-trivial.
A denominator formula for positive levels for $W(\fg^+)$
coincides with the one for
Borcherds superalgebras \cite{Cederwall:2015oua}; we do not yet have
such a formula for $S(\fg^+)$ in situations where the ideal $K$ is
non-trivial. Neither do we have a denominator formula for negative
levels. In situations described in the end of Section \ref{affinesection}, where an invariant bilinear form exists, the
negative level generators can be deduced from the positive ones. This invariant bilinear form is interesting for many other reasons too,
and needs to be better understood.

\section*{Acknowledgements}

We would like to thank V.~Kac for discussions, and in particular for calling our attention to the multiplicity formula (\ref{PRBFormula}) in ref.~\cite{ParthasarathyEtAl}.
This research is supported by the
Swedish Research Council, project no. 2015-04268.

\bibliographystyle{utphysmod2}



\begin{thebibliography}{10}

\bibitem{CederwallPalmkvistTHAII}
M.~Cederwall and J.~Palmkvist,  {\em {Tensor hierarchy algebras and extended
  geometry II: Gauge structure and dynamics}},
\href{http://www.arXiv.org/abs/yymm.nnnnn}{{\tt yymm.nnnnn}}.

\bibitem{Kac77B}
V.~G. Kac,  {\em {L}ie superalgebras}, Adv. Math. {\bf 26}, 8--96 (1977).

\bibitem{Carbone:2018xqq}
L.~Carbone, M.~Cederwall and J.~Palmkvist,  {\em {Generators and relations for
  Lie superalgebras of Cartan type}}, J. Phys. {\bf A52}, 055203 (2019)
[\href{http://www.arXiv.org/abs/1802.05767}{{\tt 1802.05767}}].

\bibitem{Palmkvist:2013vya}
J.~Palmkvist,  {\em {The tensor hierarchy algebra}}, J. Math. Phys. {\bf 55},
  011701 (2014)
[\href{http://www.arXiv.org/abs/1305.0018}{{\tt 1305.0018}}].

\bibitem{Greitz:2013pua}
J.~Greitz, P.~Howe and J.~Palmkvist,  {\em {The tensor hierarchy simplified}},
Class.\ Quant.\ Grav.\  {\bf 31}, 087001 (2014) 
\href{http://www.arXiv.org/abs/1308.4972}{{\tt 1308.4972}}.

\bibitem{Bossard:2017wxl}
G.~Bossard, A.~Kleinschmidt, J.~Palmkvist, C.~N. Pope and E.~Sezgin,  {\em
  {Beyond $E_{11}$}}, JHEP {\bf 05}, 020 (2017)
[\href{http://www.arXiv.org/abs/1703.01305}{{\tt 1703.01305}}].

\bibitem{Bossard:2017aae}
G.~Bossard, M.~Cederwall, A.~Kleinschmidt, J.~Palmkvist and H.~Samtleben,  {\em
  {Generalized diffeomorphisms for $E_9$}}, Phys. Rev. {\bf D96}, 106022 (2017)
[\href{http://www.arXiv.org/abs/1708.08936}{{\tt 1708.08936}}].

\bibitem{Cederwall:2018aab}
M.~Cederwall and J.~Palmkvist,  {\em {$L_{\infty }$ algebras for extended
  geometry from Borcherds superalgebras}}, Commun. Math. Phys. {\bf 369},
  721--760 (2019)
[\href{http://www.arXiv.org/abs/1804.04377}{{\tt 1804.04377}}].

\bibitem{Bossard:2019ksx}
G.~Bossard, A.~Kleinschmidt and E.~Sezgin,  {\em {On supersymmetric E$_{11}$
  exceptional field theory}},
\href{http://www.arXiv.org/abs/1907.02080}{{\tt 1907.02080}}.

\bibitem{Cederwall:2017fjm}
M.~Cederwall and J.~Palmkvist,  {\em {Extended geometries}}, JHEP {\bf 02}, 071
  (2018)
[\href{http://www.arXiv.org/abs/1711.07694}{{\tt 1711.07694}}].

\bibitem{ParthasarathyEtAl}
K.~Parthasarathy, R.~Ranga~Rao and V.~Varadarajan,  {\em {Representations of
  complex semi-simple Lie groups and Lie algebras}}, Ann. Math. {\bf 85}, 383
  (1967).

\bibitem{Dobrev:1985qz}
V.~K. Dobrev and V.~B. Petkova,  {\em {Group theoretical approach to extended
  conformal supersymmetry: Function space realizations and invariant
  differential operators}}, Fortsch. Phys. {\bf 35}, 537
(1987).

\bibitem{Palmkvist:2015dea}
J.~Palmkvist,  {\em {Exceptional geometry and Borcherds superalgebras}}, JHEP
  {\bf 11}, 032 (2015)
[\href{http://www.arXiv.org/abs/1507.08828}{{\tt 1507.08828}}].

\bibitem{Kantor-graded}
I.~L. Kantor,  {\em Graded {L}ie algebras}, Trudy Sem. Vect. Tens. Anal. {\bf
  15}, 227--266 (1970).

\bibitem{Palmkvist:2009qq}
J.~Palmkvist,  {\em {Three-algebras, triple systems and 3-graded Lie
  superalgebras}}, J.\ Phys.\ A {\bf A43}, 015205 (2010)
[\href{http://www.arXiv.org/abs/0905.2468}{{\tt 0905.2468}}].

\bibitem{Berman:2012vc}
D.~S. Berman, M.~Cederwall, A.~Kleinschmidt and D.~C. Thompson,  {\em {The
  gauge structure of generalised diffeomorphisms}}, JHEP {\bf 01}, 064 (2013)
[\href{http://www.arXiv.org/abs/1208.5884}{{\tt 1208.5884}}].

\bibitem{Cederwall:2015oua}
M.~Cederwall and J.~Palmkvist,  {\em {Superalgebras, constraints and partition
  functions}}, JHEP {\bf 08}, 036 (2015)
[\href{http://www.arXiv.org/abs/1503.06215}{{\tt 1503.06215}}].

\bibitem{Howe:2015hpa}
P.~Howe and J.~Palmkvist,  {\em {Forms and algebras in (half-)maximal
  supergravity theories}}, JHEP {\bf 05}, 032 (2015)
[\href{http://www.arXiv.org/abs/1503.00015}{{\tt 1503.00015}}].

\bibitem{Kleinschmidt:2013em}
A.~Kleinschmidt and J.~Palmkvist,  {\em {Oxidizing Borcherds symmetries}}, JHEP
  {\bf 1303}, 044 (2013)
[\href{http://www.arXiv.org/abs/1301.1346}{{\tt 1301.1346}}].

\end{thebibliography}

\providecommand{\href}[2]{#2}\begingroup\raggedright\endgroup

\end{document}